\documentclass[12pt]{iopart}

\usepackage{iopams,amssymb,graphicx,hyperref}  
\newcommand{\beq}{\begin{equation}}
\newcommand{\eeq}{\end{equation}}
\newcommand{\be}{\begin{equation}}
\newcommand{\ee}{\end{equation}}
\newcommand{\bea}{\begin{eqnarray}}
\newcommand{\eea}{\end{eqnarray}}
\newcommand{\lrf}[2]{\left(\frac{#1}{#2}\right)}

\newcommand{\nnmb}{\nonumber}

\newcommand{\gev}{\, {\rm GeV}}

\newcommand{\ecm}{\, e\;{\rm cm}}

\begin{document}

\noindent{\vbox{ \hbox{NPAC-12-08} }}


\title[Electroweak baryogenesis]{Electroweak baryogenesis}

\author{David E. Morrissey$^1$ and Michael J. Ramsey-Musolf$^{2,3}$}

\address{
$^1$ TRIUMF, 4004 Wesbrook Mall, Vancouver, BC V6T 2A3, Canada}
\address{
$^2$ Department of Physics, University of Wisconsin, Madison, WI 53705, USA}
\address{
$^3$ Kellogg Radiation Laboratory, California Institute of Technology, Pasadena, CA 91125 USA}
\ead{dmorri@triumf.ca, mjrm@physics.wisc.edu}

\begin{abstract}

Electroweak baryogenesis (EWBG) remains a theoretically attractive and experimentally testable scenario for explaining the cosmic baryon asymmetry. We review recent
progress in computations of the baryon asymmetry within this framework and  discuss their phenomenological consequences. We pay particular attention to methods for
analyzing the electroweak phase transition and calculating CP-violating asymmetries, the development of Standard Model extensions that may provide the necessary ingredients
for EWBG, and searches for corresponding signatures at the high energy, intensity, and cosmological frontiers. 

\end{abstract}

\maketitle


\section{Introduction}

  Electroweak baryogenesis~(EWBG) is one of the most attractive and promising 
ways to account for the observed baryon asymmetry of the Universe~\cite{Kuzmin:1985mm,Shaposhnikov:1986jp,Shaposhnikov:1987tw}.  
As its name suggests, EWBG refers to any mechanism that produces an asymmetry 
in the density of baryons during the electroweak phase transition.  
While many specific realizations of EWBG have been proposed,
they all have many features in common, and in this review we attempt 
to describe them in a unified way.\footnote{See Refs.~\cite{Cohen:1993nk,Trodden:1998ym,Riotto:1998bt,Riotto:1999yt,Quiros:1999jp,Dine:2003ax,Cline:2006ts} for previous reviews of EWBG.}  

  The initial conditions assumed for EWBG are a hot, radiation-dominated 
early Universe containing zero net baryon charge in which 
the full $SU(2)_L\times U(1)_Y$ electroweak symmetry 
is manifest~\cite{Kirzhnits:1972iw,Kirzhnits:1972ut,Dolan:1973qd,Dolan:1973qd,Weinberg:1974hy}.  
As the Universe cools to temperatures below the electroweak scale,
$T \lesssim 100\gev$, the Higgs field settles into a vacuum state
that spontaneously breaks the electroweak symmetry down
to its $U(1)_{em}$ subgroup.  It is during this phase transition
that EWBG takes place.

\begin{figure}[ttt]
\begin{center}
  \includegraphics[width=0.47\textwidth]{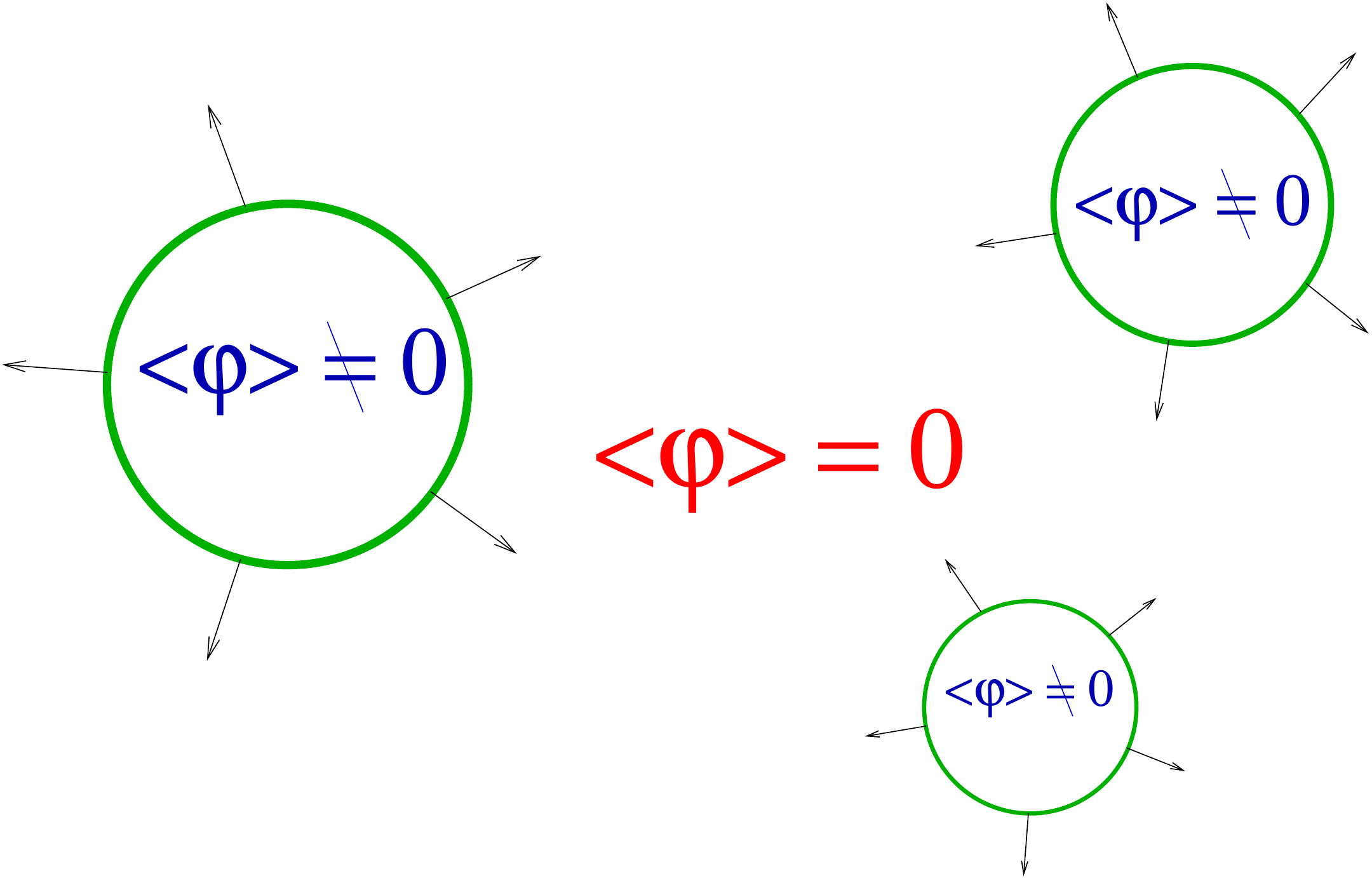}
\end{center}
\caption{\label{fig:bubble-a}
Expanding bubbles of electroweak-broken phase within the surrounding
plasma in the electroweak-symmetric phase.}
\end{figure}

  Successful EWBG requires a first-order electroweak phase transition.
Such a transition proceeds when bubbles of the broken phase nucleate 
within the surrounding plasma in the symmetric phase.  We illustrate
this process in Fig.~\ref{fig:bubble-a}.  
These bubbles expand, collide, and coalesce until only the broken 
phase remains.  

\begin{figure}[ttt]
\begin{center}
  \includegraphics[width=0.5\textwidth]{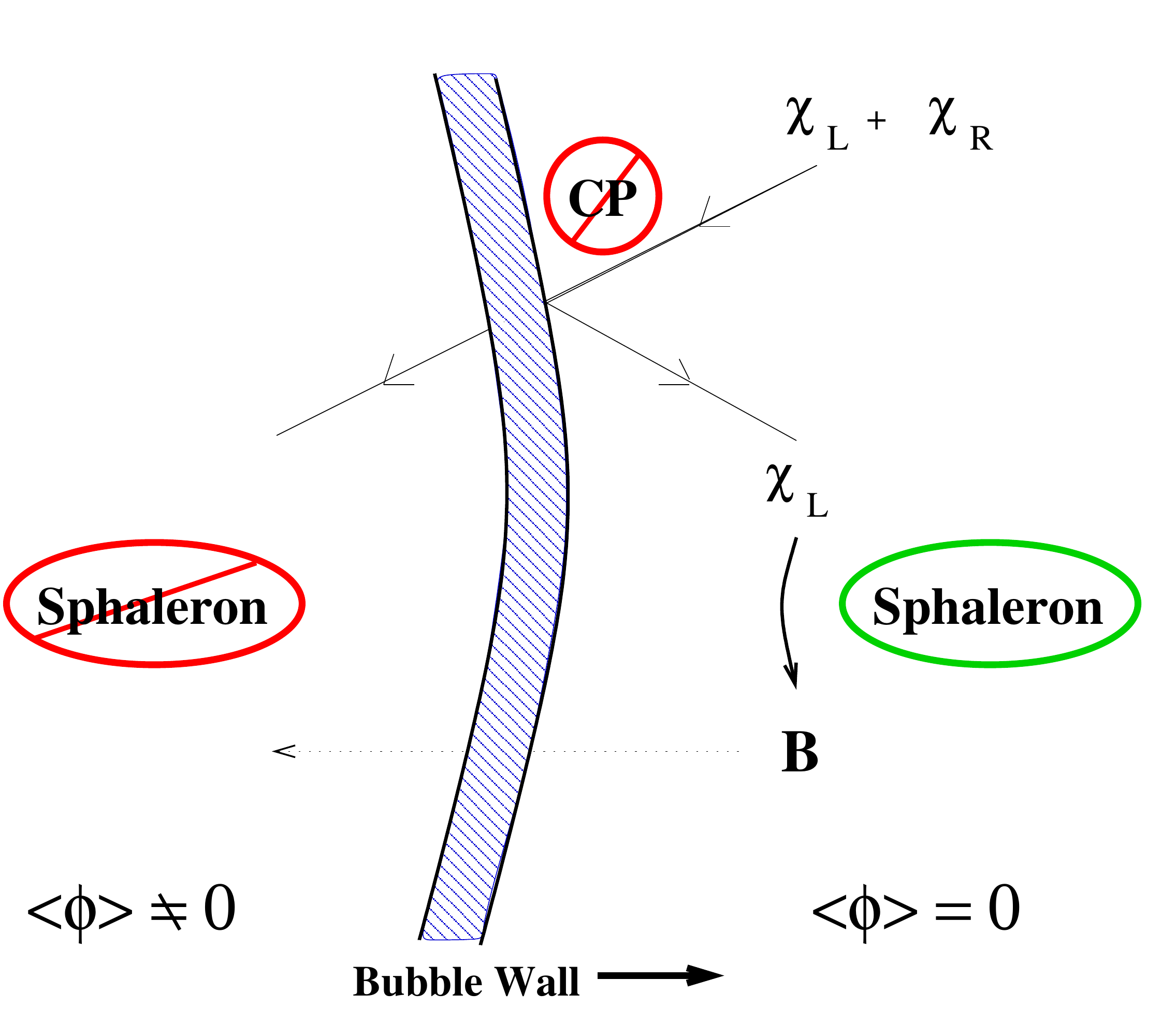}
\end{center}
\caption{\label{fig:bubble-b}
Baryon production in front of the bubble walls.}
\end{figure}

  Baryon creation in EWBG takes place in the vicinity of the expanding
bubble walls.  The process can be divided into three steps~\cite{Cohen:1993nk}:
\begin{itemize}
\item[1.] Particles in the plasma scatter with the bubble walls.  
If the underlying theory contains CP violation, this scattering 
can generate CP (and C) asymmetries in particle number densities in front 
of the bubble wall.
\item[2.] These asymmetries diffuse into the symmetric phase ahead of
the bubble wall, where they bias electroweak sphaleron 
transitions~\cite{Manton:1983nd,Klinkhamer:1984di} to produce more baryons 
than antibaryons.  
\item[3.] Some of the net baryon charge created outside the bubble wall 
is swept up by the expanding wall into the broken phase.  
In this phase, the rate of sphaleron transitions is strongly suppressed, 
and can be small enough to avoid washing out the baryons created in
first two steps.
\end{itemize}  
We illustrate these three steps in Fig.~\ref{fig:bubble-b}.

  These EWBG steps satisfy explicitly the three
Sakharov conditions for baryon creation~\cite{Sakharov:1967dj}.  
First, departure from thermodynamic equilibrium is induced by 
the passage of the rapidly-expanding bubble walls through the 
cosmological plasma.  Second, violation of baryon number comes from
the rapid sphaleron transitions in the symmetric phase.  And third, both
C- and CP-violating scattering processes are needed at the phase
boundaries to create the particle number asymmetries that bias the sphalerons
to create more baryons than antibaryons.

  All the ingredients required for EWBG are contained in the SM.
Unfortunately, EWBG is unable to explain the observed baryon asymmetry
within the SM alone.  The first impediment is that the SM electroweak phase
transition is first-order only if the mass of the Higgs boson lies 
below $m_h \lesssim 70\,\gev$~\cite{Bochkarev:1987wf,Kajantie:1995kf}.
This is much less than the current experimental lower bound of 
$m_h > 115.5\,\gev$~\cite{Barate:2003sz,:2012si}.
Even if the phase transition were first order, the CP violation
induced by the CKM phase does not appear to be sufficient to generate
large enough chiral asymmetries~\cite{Gavela:1993ts,Huet:1994jb,Gavela:1994dt}.

  Therefore an essential feature of all viable realizations of EWBG is  
new physics beyond the Standard Model~(SM).  This beyond the SM (BSM) physics
must couple to the SM with at least a moderate strength, and it must
be abundant in the thermal plasma at the time of the electroweak phase
transition.  Together, these two conditions
imply the existence of new particles with masses not too far above 
the electroweak scale and direct couplings to the SM.  Thus, a generic
prediction of EWBG is that new phenomena should be discovered in upcoming
collider and precision experiments.  It is this property that sets
EWBG apart from many other mechanisms of baryon creation.

Because of the prospects for experimental probes of EWBG, it is particularly
important to achieve the most robust theoretical predictions for the baryon asymmetry 
within this framework as well as for the associated phenomenological implications within 
specific BSM scenarios. Consequently, we review both progress in  developing the
theoretical machinery used for computations of the baryon asymmetry as well as developments
on the phenomenological front. The former entail a mix of non-perturbative Monte Carlo studies and 
various perturbative approximations. Work on the phenomenological side includes applications to
specific BSM scenarios, such as the Minimal Supersymmetric Standard Model (MSSM), and the delineation
of consequences for collider studies, low-energy probes of CP-violation, and astrophysical observations. 

  The plan for this review is as follows.  In Section~\ref{sec:ewpt} 
we discuss the electroweak phase transition in greater depth, 
concentrating on its strength and other characteristics.  
Next, in Section~\ref{sec:cpbg} we describe in more detail the creation of
asymmetries in the CP and baryon charges during the phase transition. 
Some of the ways the new ingredients required for EWBG can  
be studied in the laboratory are studied in Section~\ref{sec:test}.  
Finally, Section~\ref{sec:conc} is reserved for our conclusions.

\section{The electroweak phase transition
\label{sec:ewpt}}

  Baryon creation in EWBG is closely tied to the dynamics of the electroweak
phase transition~(EWPT).  In this transition, the thermal plasma goes from a
\emph{symmetric} state in which the full $SU(2)_L\times U(1)_Y$ 
gauge invariance is manifest to a \emph{broken} one where only 
the $U(1)_{em}$ electroweak subgroup 
remains~\cite{Kirzhnits:1972iw,Dolan:1973qd,Weinberg:1974hy}.  
As discussed above, the transition must be first-order and proceed
through the nucleation of bubbles of the broken phase.
In this section we will discuss the dynamics of this phase
transition and describe the role it plays in EWBG.

  The transition from symmetric to broken phase in the SM 
can be characterized by the vacuum expectation value~(VEV) of the 
Higgs field $H \equiv (H^+,\,H^0)^T$ that transforms as 
$(\mathbf{1},\mathbf{2},1/2) $ under $SU(3)_c\times SU(2)_L\times U(1)_Y$.  
A field basis can always be chosen such that only the real component
of $H^0$ develops a non-zero expectation value.  Thus, we will write
\beq
\phi/\sqrt{2} \equiv \langle H^0\rangle \ .
\eeq
The symmetric phase corresponds to $\phi=0$ and the broken phase
to $\phi \neq 0$.  Note that (in unitary gauge) the masses of 
the $W^{\pm}$ and $Z^0$ weak vector bosons and the fermions are
proportional to $\phi$.  

The features of this transition that are most relevant for EWBG are (a) its character (first order, second order, cross over); (b) the critical temperature $T_c$ and the bubble nucleation temperature $T_n$ that describe when it occurs; (c) the sphaleron transition rate $\Gamma_\mathrm{sph}$ that governs the rate of baryon number generation and washout; and (d) the bubble nucleation rate.  These features have been studied using a broad range of theoretic tools.

The most robust computations of many of these quantities are performed
using non-perturbative, Monte Carlo methods. However, given the level of effort required to perform such studies, they have only been applied to a few specific theories of electroweak symmetry breaking.  Instead, perturbative methods have been used much more frequently to study the dynamics of the EWPT in a broad range of BSM scenarios.
Perturbative analyses can also provide helpful insight into some aspects of phase transition dynamics that be may be less accessible with Monte Carlo methods. It should be emphasized, however, that the application of perturbation theory to EWPT physics is fraught with uncertainties as well as the potential for ambiguities. In the SM, for example, one often finds the transition temperature computed perturbatively to be significantly lower than the value obtained from Monte Carlo studies for a given value of the Higgs boson mass.  Nevertheless, given the widespread use of perturbation theory, we begin by reviewing this approach, first laying out the conventional treatment and then commenting on various difficulties.  We subsequently review salient features of the non-perturbative analyses.

\subsection{Perturbative analyses}

  In a perturbative analysis of the electroweak phase transition, the
central object is the (renormalized) finite-temperature effective potential.
This quantity coincides with the free energy of the cosmological
plasma~\cite{Quiros:1999jp,Kapusta:2006pm,lebellac,Das:1997gg},
provided it is reasonably close to thermodynamic equilibrium.
The key feature of the effective potential is that the expectation 
value $\phi$ of the Higgs field is that which minimises its value.  

  To one-loop order, the effective potential is given by~\cite{Quiros:1999jp}
\beq
V_{eff}(\phi,T) = V_{0}(\phi)+V_{1}(\phi)
+ \Delta V_{1}^{(T)}(\phi,T) \ ,
\eeq
where $V_0=-m^2\phi^2/2+\lambda\phi^4/4$ is the tree-level potential, $V_1(\phi)^{(0)}$ is the one-loop
effective potential at $T=0$, and $V_{1}^{(T)}(\phi,T)$ contains
the leading thermal corrections.  

  The expression for $V_1(\phi)$ is
well-known~\cite{Coleman:1973jx}:
\beq
V_1(\phi) = \sum_i\frac{n_i(-1)^{2s_i}}{4(4\pi)^2}m_i^4(\phi)
\left[\ln\lrf{m_i^2(\phi)}{\mu^2}-\mathcal{C}_i\right] \ ,
\eeq
where the sum $i$ runs over all particles in the theory,
each with $n_i$ degrees of freedom, field-dependent mass $m_i(\phi)$,
and spin.  Furthermore, we have assumed a mass-independent renormalization
with $\mu$ as the renormalization scale and the $\mathcal{C}_i$ 
are scheme-dependent constants.  
Choosing $\mu \sim \max\{m_i(\phi)\}$ optimizes the perturbative expansion.
Let us also mention that Fadeev-Popov ghosts are massless and decouple 
if we work in the Landau gauge ($\xi = 0$), in which case $n_i=3$ should
be used for each vector boson and contributions from Goldstone
bosons included as well.

  The thermal corrections are given at one-loop order by~\cite{Quiros:1999jp}
\beq
\Delta V_{1}^{(T)}(\phi,T) =
\sum_{i=boson}n_i\frac{T^4}{2\pi^2}J_b\!\lrf{m^2_i}{T^2}
-\sum_{j=fermion}n_j\frac{T^4}{2\pi^2}J_f\!\lrf{m^2_j}{T^2}
\eeq
where $J_b$ and $J_f$ are loop functions.  For small arguments, $x\ll 1$,
they have the expansions~\cite{Anderson:1991zb}
\bea
J_b(x^2) &=& -\frac{\pi^4}{45}+\frac{\pi^2}{12}x^2-\frac{\pi}{6}x^3
-\frac{1}{32}x^4\ln(x^2/a_b) + \mathcal{O}(x^3)
\label{eq:jb}\\
J_f(x^2) &=& -\frac{7\pi^4}{360}-\frac{\pi^2}{24}x^2
-\frac{1}{32}x^4\ln(x^2/a_f) + \mathcal{O}(x^3)
\label{eq:jf}
\eea
with $\ln(a_b) \simeq 5.4076$ and $\ln(a_f) \simeq 2.6351$.
At large arguments, $x\gg 1$, both loop functions 
reduce to~\cite{Anderson:1991zb}
\beq
J_b(x^2) \simeq J_f(x^2) = \lrf{x}{2\pi}^{3/2}e^{-x}\left(
1+\frac{15}{8x} + \mathcal{O}({x^{-2}})\right) \ .
\eeq
This form shows the familiar Boltzmann suppression of particles
much heavier than the temperature.

  To illustrate the effect of thermal corrections on the Higgs
potential, it is helpful to write the potential in a simplified approximate
form using the high-temperature expansions of Eqs.(\ref{eq:jb},\ref{eq:jf}),
noting that heavy particles ($m\gg T$) decouple quickly.
This yields~\cite{Anderson:1991zb}
\beq
V_{eff}(\phi,T) \simeq D(T^2-T_0^2)\phi^2 - {E}T\phi^3
+\frac{\bar{\lambda}}{4}\phi^4 \ ,
\label{eq:vsimple}
\eeq
where $D$ and $\bar{\lambda}$ are slowly-varying functions of $T$
(but not $\phi$).  

\begin{figure}[ttt]
\begin{center}
  \includegraphics[width=0.5\textwidth]{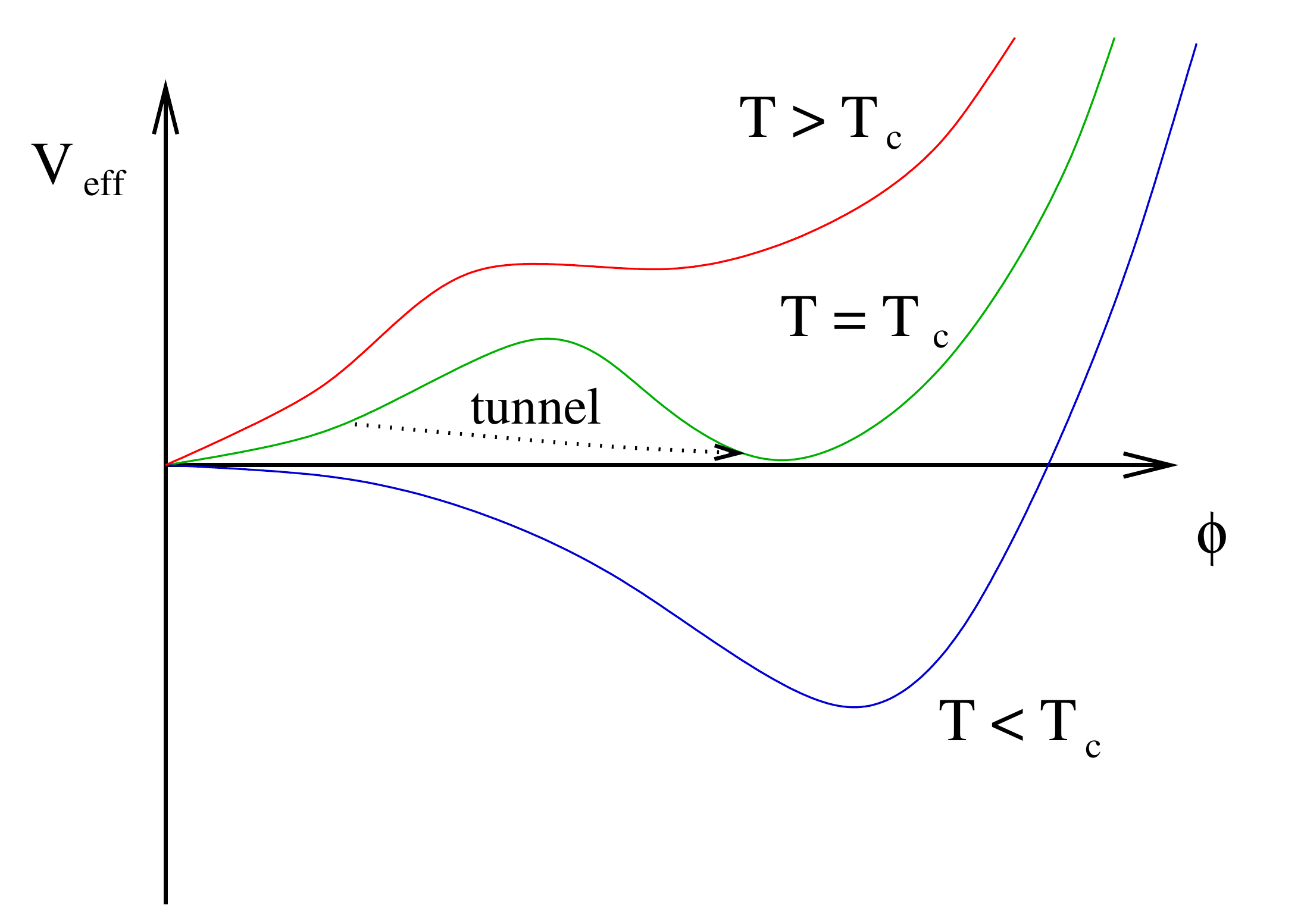}
\end{center}
\caption{\label{fig:veff}
Schematic temperature dependence of the effective potential.}
\end{figure}

  In the limit of $E=0$ in Eq.(\ref{eq:vsimple}) the phase 
transition is second order, 
with a transition temperature of $T = T_0$ and a Higgs 
expectation value for $T<T_0$ of
\beq
\phi = T_0\sqrt{\frac{2D}{\bar{\lambda}}(1-T^2/T_0^2)} \ .
\eeq
For non-zero $E$ in Eq.(\ref{eq:vsimple}), the phase transition 
becomes first-order.  Starting from $T\gg T_0$, a second
minimum away from the origin develops when $T=T_1$ with
\beq
T_1 = T_0\,\sqrt{\frac{8\bar{\lambda}D}{8\bar{\lambda}DT_0^2-9E^2}} \ ,
\eeq
where the temperature-dependent coefficients $D$ and $\bar{\lambda}$ are
to be evaluated at $T=T_1$.
This second symmetry-breaking minimum becomes degenerate with
the origin at the \emph{critical temperature} $T_c$, and becomes
deeper at lower temperature, as illustrated in Fig.\ref{fig:veff}.  
The degree to which the phase transition
is first order is typically characterized by $\phi_c/T_c$, 
where $\phi_c$ is the location of the minimum at $T_c$.  
In terms of the parameters in the potential, it is
\beq
\label{eq:phitc1}
\frac{\phi_c}{T_c} = \frac{2E}{\bar{\lambda}} \ .
\eeq
Some time after the temperature falls below $T_c$, regions of
the cosmological plasma tunnel to the deeper broken minimum
and the phase transition proceeds by the nucleation of
bubbles.

  Before discussing the dynamics of such a first-order phase transition,
it is worth examining the validity of the perturbative expansion
outlined above.  This expansion is known to break down at very high
temperatures when the thermal loop expansion
parameter $g^2T^2/m^2(\phi)$ becomes large~\cite{Weinberg:1974hy,Gross:1980br}
(where $g^2$ is the coupling entering in the loop).  
Indeed, we saw above that the
leading thermal corrections, which are only generated by loops in this
formalism, completely change the vacuum structure of the theory
and induce a restoration of symmetry at very high temperatures.
%
%
The breakdown of the perturbative expansion can be postponed 
by resumming the most dangerous thermal corrections by incorporating
thermal mass corrections in the propagators.  
The net result of such a \emph{daisy resummation} is to 
generate an additional term  in the effective potential~\cite{Parwani:1991gq}:
\beq
V_1^{(daisy)} = -\frac{T}{12\pi}\sum_{\{b\}'}n_b
\left[\overline{m}_b^3(\phi,T)-m^3_b(\phi)\right]^{3/2} \ ,
\label{eq:daisy}
\eeq
where the sum runs only over scalars and longitudinal vectors,
and $\overline{m}^2$ is the field-dependent thermal squared mass:
\beq
\overline{m}^2(\phi) = m^2(\phi) + \Pi(T) \ ,
\eeq
with $\Pi(T) \propto T^2$ the thermal contribution to the mass.

  The daisy correction is particularly important for a first-order
transition because it affects primarily the crucial cubic term.
For example, suppose the contribution to the cubic term comes 
from a scalar with a zero-temperature mass of $m^2(\phi) = g\phi^2$ 
with a thermal correction of $\Pi(T) = \kappa\,T^2$.
The would-be cubic term becomes
\beq
\Delta E\phi^3 = \frac{1}{12\pi}g^{3/2}\phi^3 \to 
\frac{1}{12\pi}\left[g\phi^2+\kappa\,T^2\right]^{3/2}
\eeq
When $\Pi(T)$ is large relative to $m^2(\phi)$, this corrected
expression ceases to behave as a cubic in $\phi$ and the phase
transition might no longer be first-order.

  When the electroweak phase transition is first-order, it proceeds
by the nucleation of bubbles of the broken phase within the surrounding
plasma of symmetric phase.  Bubble nucleation is governed by thermal
tunnelling~\cite{Linde:1977mm} from the local minimum at $\phi = 0$
to a deeper minimum at $\phi \neq 0$.
In nucleating a bubble there is a competition between the decrease
in free energy, proportional to bubble volume, with the increase 
due to the tension of the wall, proportional to bubble area.  
As such, there is a minimum radius for which a bubble can grow after
it is formed, and this limits the tunnelling rate.  Bubble formation 
and growth only begins in earnest when this rate exceeds the Hubble rate, 
which occurs at some temperature $T_n < T_c$, called the 
\emph{nucleation temperature}.  
Once a sufficiently large bubble is formed, it expands until it
collides with other bubbles and the Universe is filled with the broken 
phase.   The typical profile and expansion rate of a bubble wall can 
be computed from the effective potential~\cite{Anderson:1991zb,Dine:1992wr,Moore:2000jw}, taking into account
frictional effects from scattering with surrounding particles in the
plasma~\cite{Moore:2000wx,Megevand:2009gh,Espinosa:2010hh}.  
As we will see in the next section, baryon creation 
processes are very sensitive to the speed and profile of the walls.

  A first-order electroweak phase transition is not sufficient  
for successful EWBG.  The transition must also be 
\emph{strongly first-order}.  
Within the context of perturbation theory calculations, 
the quantitative condition for a strongly first-order phase 
transition has typically been taken to be
\beq
\frac{\phi_c}{T_c} \gtrsim 1 \ .
\label{eq:firstorder}
\eeq
This ratio approximates a factor that appears in the rate for sphaleron
transitions in the broken phase within the bubble walls.  When this
condition is not met, these transitions will wash out the baryons
created by EWBG.  As we will describe below, 
the necessity of a very strong phase transition is one of the reasons 
why EWBG does not work in the SM, and is therefore a strong motivator 
of new physics beyond the SM.

  The condition of Eq.(\ref{eq:firstorder}) is a  frequently-applied
approximation, but a precision calculation of the baryon abundance from EWBG 
requires a more careful analysis.  For one, the relevant temperature
for the dynamics is the slightly lower bubble nucleation temperature 
rather than the critical temperature, 
although it is often the case that $T_n \simeq T_c$~\cite{Carena:2008rt,Carena:2008vj}.  
A more serious problem is the lack of gauge invariance.
In particular, it is well-known that the vacuum expectation value 
of the one-loop effective potential $\phi$ at any temperature
is gauge-dependent~\cite{Dolan:1974gu,Nielsen:1975fs,Fukuda:1975di,Fischler:1974ue,Baacke:1993aj,Baacke:1993aj,Baacke:1994ix,Laine:1994zq,Metaxas:1995ab,Boyanovsky:1996dc,Patel:2011th,Wainwright:2011qy}, 
so that the ratio on the left-hand side of Eq.(\ref{eq:firstorder}) 
is not a well-defined physical quantity.   Moreover, the procedure 
outline above to determine $T_c$ perturbatively also introduces 
a spurious gauge-dependence, as one may observe by expressing the right
side of Eq.(\ref{eq:phitc1}) in an arbitrary gauge~\cite{Patel:2011th}:
\beq
\label{eq:phitc2}
\frac{\phi_c}{T_c}=\frac{2 E}{\bar\lambda} = \frac{3-\xi^{3/2}}{48\pi\lambda}\ \left[2 g_2^2+(g_1^2+g_2^2)^{3/2}\right]+\cdots\ \ \ ,
\eeq
where the additional terms are $\xi$-dependent contributions associated with the one-loop corrections to the Higgs quartic self-coupling. As indicated above, the conventional analyses have been
performed in Landau gauge, even though a small change in the choice of gauge parameter can significantly alter the ratio $\phi_c/T_c$. 

Obtaining a gauge-invariant {\em baryon number preservation criterion}~(BNPC) requires several modifications of the na\"ive perturbative treatment described above:
\begin{itemize}
\item[1.] Determine $T_c$ (or $T_n$) in a gauge-invariant manner by following the evolution of $V_\mathrm{eff}(\phi,T)$ and consistently implementing the so-called Nielsen identities~\cite{Nielsen:1975fs,Fukuda:1975di}. One procedure for doing so entails carrying out an $\hbar$-expansion of $V_\mathrm{eff}(\phi,T)$, as outlined in Refs.~\cite{Fukuda:1975di,Laine:1994zq,Patel:2011th}. A generalization of this procedure also allows one to approximate the full daisy resummation in a gauge-invariant way and to reproduce the trends for $T_c$ found in nonperturbative calculations~\cite{Patel:2011th}. 
\item[2.] Perform a gauge-invariant computation of the energy of the sphaleron configuration, $E_\mathrm{sph}$. 
In perturbation theory, it is possible to do so in the broken phase by working with the high-temperature effective theory in which zero-temperature masses are replaced by their (gauge-invariant) Debye masses. The energy $E_\mathrm{sph}$ then depends on a gauge-invariant scale ${\bar v}(T)$ that is not the same as $\phi(T)$, and the fluctuation determinant $\kappa$ that characterizes the leading quadratic corrections to the sphaleron action~\cite{Arnold:1987mh,Carson:1990jm}. 
\item[3.] Compute the baryon density $n_B$ at the end of the EWPT, 
corresponding to a time delay of $\Delta t_\mathrm{EW}$ after its onset, 
and compare with the initial density resulting from the CPV transport dynamics described in Section~\ref{sec:cpbg} below. 
The resulting ratio is called the \lq\lq washout factor"
\beq
\label{eq:washout1}
S=\frac{n_B(\Delta t_\mathrm{EW})}{n_B(0)}\ \ \ .
\eeq
For the baryon asymmetry created by EWBG to be preserved, the washout
factor $S$ must not be too small.
\end{itemize}
Rewriting the washout factor in terms of $X$, defined according to
\beq
\label{eq:washout2}
S > e^{-X} \ ,
\eeq
the quantitative BNPC is~\cite{Patel:2011th}:
\beq
\label{eq:bnpc}
\frac{4\pi B}{g}\ \frac{{\bar v}(T_c)}{T_c} - 7 \ln\frac{{\bar v}(T_c)}{T_c} > -\ln X
-\ln \left( \frac{\Delta t_\mathrm{EW}}{t_H}  \right) + \ln\mathcal{Q}\mathcal{F} +\hbar\ln\kappa \ .
\eeq
Here $B$ parameterizes the relationship between the scale ${\bar v}(T)$ and the sphaleron energy
~\cite{Manton:1983nd,Brihaye:1993ud}
\beq
E_\mathrm{sph}(T) \simeq B\,\frac{2m_W}{\alpha_w}\,\lrf{{\bar v}(T)}{v(0)} \ ,
\label{eq:esphal}
\eeq
where $B$ is a constant of order unity that depends on the mass
of the Higgs boson,  $v(0)=174\,\gev$ is the value of the Higgs VEV at $T=0$, and $\alpha_w$ is the weak coupling. The other quantities 
in Eq.(\ref{eq:bnpc}) include: the Hubble time $t_H$; a quantity $\mathcal{Q}$
characterizing the contribution of sphaleron zero modes; 
a function $\mathcal{F}$ that characterizes the dependence of the unstable mode 
of the sphaleron on ${\bar v}(T)$; and a factor $\kappa$ accounting for
fluctuations that are not zero modes. 
The appearance of the logarithms in Eq.(\ref{eq:bnpc}) and the 
dependence on $\Delta t_\mathrm{EW}$ result from integrating the
baryon number evolution equation~\cite{Bochkarev:1987wf,Bodeker:1999gx}
\beq
\frac{dn_B}{dt}\simeq -\frac{13N_f}{2}\frac{\Gamma_\mathrm{ws}}{VT^3}n_B \ ,
\label{eq:washoutb}
\eeq
where $N_f$ is the number of fermion families and 
${\Gamma_\mathrm{ws}}/{VT^3} \propto \exp(-E_\mathrm{sph}/T)$
is the sphaleron rate per unit volume inside the bubble.
Qualitatively, the BNPC in Eq.(\ref{eq:bnpc}) corresponds to the
requirement that the sphaleron rate in the broken phase during
the phase transition be much slower than the Hubble expansion rate.

  The conventionally-employed condition of Eq.(\ref{eq:firstorder}) results from replacing the gauge-invariant ratio ${\bar v}(T_c)/T_c$ by the gauge-dependent one $\phi_c/T_c$ and  making specific choices for the parameters appearing in Eq.(\ref{eq:bnpc}).  In particular, it has been assumed that one may take $X=10$, corresponding to allowing the initial baryon asymmetry to be five orders of magnitude larger than what is presently observed -- an assumption that is questionable in light of recent studies of the CPV transport dynamics discussed in Section \ref{sec:cpbg}. Additional significant uncertainties are associated with the value of the fluctuation determinant $\kappa$ and the duration of the transition $\Delta t_\mathrm{EW}$. In short, even if one employs an appropriately gauge-invariant procedure to determine the degree of baryon number preservation, considerable uncertainty remains as to the precise requirement. 

Nearly all phenomenological studies carried out over the past decade or so have neglected these issues. Even if one places some trust in the use of perturbation theory to analyze EWPT dynamics relevant to EWBG, it should be clear that considerable work remains to be carried out in order to obtain a robust statement about the presence or absence of a sufficiently strong first order transition in a given BSM scenario. It may be, for example, that a given BSM scenario significantly modifies the dependence of $E_\mathrm{sph}$ on the gauge-invariant scale ${\bar v}(T_c)$, the dependence of ${\bar v}$ on $T_c$ itself, the duration of the transition, or the fluctuation determinant. From our view, then, conclusions that have been drawn as to the viability of EWBG based on existing perturbative analyses of the scalar field dynamics should be viewed as provisional at best and, in an ideal situation, revisited in light of these open theoretical issues.

\subsection{Non-perturbative studies}

  Dedicated non-perturbative numerical analyses of the EWPT have
been carried out for the SM, the Minimal Supersymmetric
Standard Model (MSSM), and the two Higgs doublet model~\cite{Kajantie:1995kf,Ambjorn:1987qu,Ambjorn:1988gf,Ambjorn:1990wn,Ambjorn:1990pu,Kajantie:1993ag,Farakos:1994xh,Kajantie:1995dw,Kajantie:1996mn,Kajantie:1996qd,Csikor:1996sp,Gurtler:1997hr,Laine:1998qk,Laine:1998jb,Csikor:1998ge,Csikor:1998eu,Moore:1998swa,Aoki:1999fi,Csikor:2000sq,Laine:2000rm}. Among the properties studied that are particularly relevant to EWBG are the critical temperature, $T_c$; the weak sphaleron rate, $\Gamma_\mathrm{sph}$ (the focus of Refs.~\cite{Ambjorn:1990wn,Ambjorn:1990pu,Moore:1998swa}), and the character of the transition (first or second order, cross over, {\em etc.}). As we discuss in Section \ref{sec:gw}, the latent heat $L$ is also critical for the amplitude of gravity wave production, and it has been studied in, for example, Ref.~\cite{Kajantie:1995kf}.  

\begin{table}
\caption{\label{tab:crit} Maximum values of the Higgs boson mass, $M_h^C$, for a first order EWPT in the SM as obtained from lattice studies.} 

\begin{indented}
\lineup
\item[]\begin{tabular}{@{}*{3}{l}}
\br                              
$\0\0\mathrm{Lattice}$&Authors&$M_h^C$ (GeV)\cr 
\mr
\0\0 4D Isotropic & \cite{Aoki:1999fi}  & $80\pm 7$\cr
\0\0 4D Anisotropic & \cite{Csikor:1998eu}  & $72.4\pm 1.7$\cr
\0\0 3D Isotropic & \cite{Laine:1998jb}  & $72.3\pm 0.7$\cr
\0\0 3D Isotropic & \cite{Gurtler:1997hr}  & $72.4\pm 0.9$\cr
\br
\end{tabular}
\end{indented}
\end{table}


The process of bubble nucleation that is the starting point for EWBG requires a first order transition. 
In the SM, the presence of a first-order transition is found to require a sufficiently light Higgs boson.  Consequently, considerable attention has been paid to the value of the maximum value of the Higgs mass for which a first-order transition exists. Representative results from lattice studies are given in Table \ref{tab:crit}. The results obtained with the three-dimensional (3D) lattices require first carrying out the procedure of dimensional reduction to a the 3D effective theory (see, {\em e.g.}, Ref.~\cite{Kajantie:1995dw}) before studying the phase transition properties of the latter using Monte Carlo methods. For either 3D or 4D studies, a criterion must be identified for determining the character of the phase transition. Among those employed are the susceptibility associated with the scalar field $\chi\propto \langle (\phi^\dag\phi-\langle\phi^\dag\phi\rangle)^2\rangle$ and correlation lengths. The scaling behavior of $\chi$ with lattice volume can be used to determine whether the transition is first order, second order, or cross over. For $M_h >> M_h^C \simeq 75\,\gev$, as implied by collider searches for the Higgs, the transition appears to be a cross over transition. 

\begin{figure}[ttt]
\begin{center}
  \includegraphics[width=0.4\textwidth]{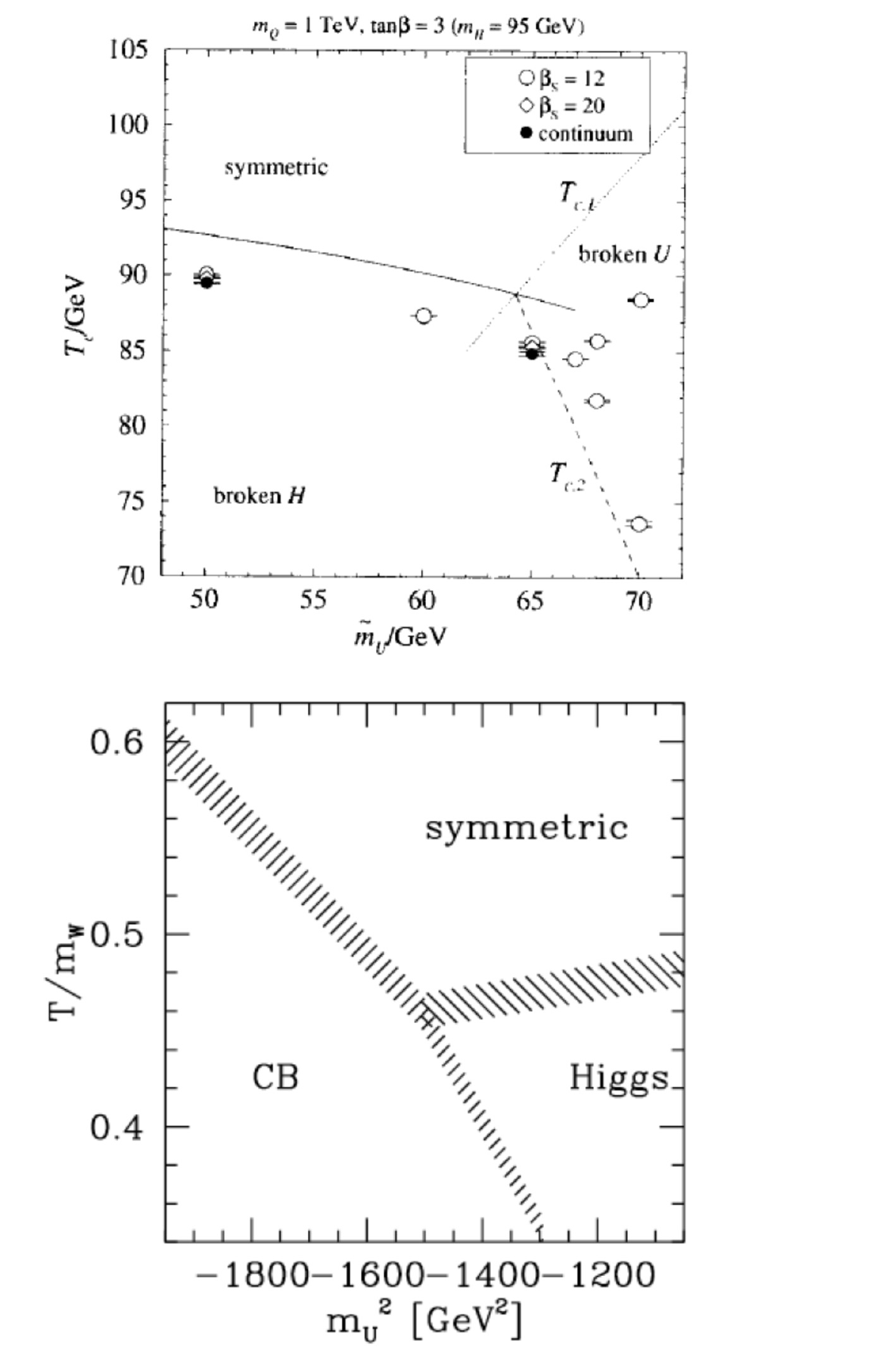}
\end{center}
\caption{\label{fig:latt}
Dependence of the critical temperature $T_c$ in the MSSM on the SUSY-breaking RH stop mass parameter $m_U$ taken from Ref.~\cite{Laine:1998qk} (top panel) and Ref.~\cite{Csikor:2000sq} (bottom panel) . Note that the contribution to the ${\tilde t}_R$ mass-squared from this parameter goes as $-{\tilde m}_U^2=m_U^2$ . The phase labeled \lq\lq CB" or \lq\lq broken U" denotes a phase in which the stop field acquires a non-zero vacuum expectation value, corresponding to a color and charge-breaking vacuum.
Top panel reprinted from M. Laine and K. Rummukainen, \href{http://www.sciencedirect.com/science/article/pii/S0550321398005306}{
Nucl. Phys. B 535, 423 (1998)} with permission from Elsevier.
Bottom panel reprinted with permission from F. Csikor, Z. Fodor, P. Hegedus, A. Jakovac, S. D. Katz and A. Piroth, 
\href{http://link.aps.org/doi/10.1103/PhysRevLett.85.932}{Phys. Rev. Lett. 85, 932 (2000)}, copyright 2000 by the American Physical Society.
}
\end{figure}

  A first-order transition can be accomplished by introducing additional scalar fields that couple to the Higgs sector. As discussed below, a particularly important example is the MSSM where the scalar top quarks couple strongly to the Higgs fields. Lattice studies imply that for sufficiently light right-handed (RH) stops, ${\tilde t}_R$, the transition can be not only first order but strongly so, and that $T_c$ decreases with the RH stop mass. The lowering of $T_c$ is important for the protection of the baryon asymmetry from washout by EW sphalerons inside the expanding bubbles, since at the onset of the transition $\Gamma_\mathrm{sph}\sim \exp (-E_\mathrm{sph}/T_c)$. Lowering $T_c$ reduces $\Gamma_\mathrm{sph}$ by both reducing the magnitude of the denominator in the exponential and increasing the sphaleron energy $E_\mathrm{sph}$.  Results for the dependence of $T_c$ on the SUSY-breaking RH stop mass parameter obtained from Refs.~\cite{Laine:1998qk,Csikor:2000sq} are shown in Fig. \ref{fig:latt}   . Note that the stop mass-squared parameter must be negative to lower $T_c$. Increasing its magnitude in this parameter space region thus lowers the overall mass of the ${\tilde t}_R$.

\subsection{Extending the SM scalar sector}
   
The need for a strongly first-order EWPT is one of the two reasons
why EWBG does not work in the SM. Indeed, the failure of EWBG in the SM (or baryogenesis of any form) is a motivation
for new physics near the electroweak scale.  Such new physics is also needed
to stabilize the electroweak scale itself, and can also account for the 
dark matter.  While a broad range of extensions of the SM have been proposed
to strengthen the EWPT to allow for EWBG, most of them fall into
two groups.  In the first group -- exemplified by the MSSM -- 
new scalars couple to the Higgs field
and enhance the cubic term in the effective potential by running in loops.  
The second group consists of scalar fields coupling to the Higgs
that develop non-trivial dynamics in the early Universe that influences
the effective Higgs potential directly.  In both cases, new light scalars
with significant couplings to the Higgs are needed, and these may lead
to observable effects at colliders.

  In the first class of new physics, in which a new scalar $X$
modifies the Higgs potential through its loop effects, 
the relevant interactions can usually be written in 
the form~\cite{Carena:2008rt,Cohen:2012zt} 
\beq
-\mathcal{L} \supset M_X^2|X|^2  + \frac{K}{6}\,|X|^4 + Q\,|X|^2|H|^2 \ .
\label{eq:hportal}
\eeq   
The third term is a \emph{Higgs portal} coupling that cannot be
forbidden by symmetries.  Assuming that $X$ does not develop
an expectation value, the mass of the physical complex scalar is 
\beq
m_X^2 = M_X^2 + \frac{Q}{2}\,\phi^2 \ .
\eeq
Applying this to the effective potential yields
\beq
\Delta V_{eff}(\phi,T) \supset 
-\frac{n_XT}{12\pi}\,\left[\Pi_X(T)+M_X^2+Q\,\phi^2/2\right]^{3/2} \ ,
\label{eq:xcube}
\eeq
where $\Pi_X(T)$ is the thermal mass of $X$ and $n_X$ is the number of
degrees of freedom.  If $Q\,\phi^2/2$ is much larger than the other
terms for $\phi \simeq \phi_c$, this correction gives a strong enhancement
of the cubic operator that drives a first-order phase transition.
If $X$ is charged under $SU(3)_c$, the contribution to the cubic is
further enhanced at two-loop order by corrections involving 
virtual gluons~\cite{Espinosa:1996qw,Cohen:2011ap}.
The net result is that a strongly first-order EWPT can obtain
for $Q\gtrsim 1$ and $M_X^2 \lesssim 0$ if $X$ is a $SU(3)_c$ triplet,
but much larger $Q$ values are needed if $X$ is 
a gauge singlet~\cite{Cohen:2011ap,Espinosa:1992hp}.  

  New scalars coupling in this way to the Higgs field have been studied 
extensively in the context of supersymmetry, and especially within
the MSSM. For early perturbative studies, see Refs.~\cite{Carena:1996wj,Carena:1997gx,Carena:1997ki}. 
In this case the $X$ scalar corresponds to the lightest scalar top quark (stop).
It is a $SU(3)_c$ triplet with electric charge $q=2/3$, and it must
be mostly right-handed to be both light and consistent with precision
electroweak data~\cite{Carena:2008rt}.   The other (mostly left-handed) 
stop must be considerably heavier to push up the mass of the 
Higgs boson~\cite{Carena:2008rt}.  
The couplings in Eq.(\ref{eq:hportal}) are fixed by supersymmetry to be 
$Q \simeq y_t^2$ and $K \simeq 4\pi\alpha_s$ (up to small corrections from 
electroweak effects and left-right stop mixing), while the $M_X^2$ 
mass term corresponds to the right-handed stop soft mass-squared.

  A first-order phase transition strong enough for EWBG is found to
be marginally consistent with the 
MSSM~\cite{Carena:2008rt,Laine:1998qk,Cline:1998hy}.  
As discussed above, in order to quench the sphaleron rate by lowering $T_c$,
a negatively-valued $M_X^2$ mass term is needed
to cancel against the thermal mass appearing 
in Eq.(\ref{eq:daisy})~\cite{Carena:2008rt}.
This has the effect of destabilizing the origin of the scalar field
space in the $X$ direction, making possible an expectation value
for the $X$ field.  Such a VEV would be disastrous since it would
imply the spontaneous breakdown of electric charge and color. 
However, a detailed computation  of the thermal evolution of the 
effective potential for $\phi$ and $X$ finds that slightly negative
values of $M_X^2$, but still large enough for EWBG, 
are phenomenologically acceptable.  In this scenario, thermal
effects can stabilize the $X$ direction more strongly than the $\phi$ direction,
and the system can evolve first to a local minimum with $\phi \neq 0$ and $X=0$.
This local minimum might not be as deep as the charge-breaking minimum
with $X\neq 0$, but it can be sufficiently long-lived to describe the
Universe we observe.  A negative $M_X^2$ also implies a very light stop mass,
which forces the other heavier stop to be extremely heavy to drive up the
mass of the lightest Higgs boson.
 
  A second way to make the electroweak phase transition more strongly
first-order is to couple the Higgs to a new scalar that develops a VEV
near the electroweak scale~\cite{Pietroni:1992in,Profumo:2007wc,Espinosa:2011ax}.  
A simple example of this is
\bea
-\mathcal{L} &\supset& m_N^2N^2 + A_NN^3 + \lambda_NN^4
\label{eq:singlet}\\
&&+(A_H\,N+\zeta_HN^2)|H|^2 + ... 
\nnmb
\eea
These interactions can allow both $H$ and $N$ to develop expectation
values, resulting in a mixing between the physical singlet and $SU(2)_L$
scalars in the theory.  When the singlet and $SU(2)_L$ mass parameters
are similar, it is convenient to track the evolution of the VEVs
in polar coordinates~\cite{Pietroni:1992in,Profumo:2007wc}:
\beq
\langle H^0\rangle = \varphi\,\cos\alpha,~~~~~ 
\langle N\rangle = \varphi\,\sin\alpha
\ .
\eeq
Applying this parametrization to Eq.(\ref{eq:singlet}), one obtains
cubic terms in the tree-level potential for $\varphi$ that can lead
to a strongly first-order electroweak phase transition.
The singlet can also strengthen the phase transition by contributing
to the loop-induced cubic term in the effective potential or by
reducing the effective Higgs quartic coupling near the critical
temperature~\cite{Profumo:2007wc}.
Similar effects have been found in gauge extensions of the
SM~\cite{Kang:2004pp} as well as theories with two 
or more $SU(2)_L$ doublets~\cite{Laine:2000rm,Turok:1990zg,Cline:2011mm,Borah:2012pu}.

  When the characteristic mass scale of the singlet sector is significantly 
larger than the $SU(2)_L$ part, the singlet can be integrated out of the theory.
This produces effective Higgs interactions of the form
\beq
-\mathcal{L} \supset -\mu^2|H|^2+\lambda|H|^4+\frac{1}{\Lambda_N^2}|H|^6+\ldots
\eeq
where $\Lambda_N$ characterises the mass scale of the singlet sector.
For values not too large, $\Lambda_N\lesssim 1000\,\gev$, the new sextic 
term can also drive a strongly first-order 
electroweak phase transition~\cite{Delaunay:2007wb,Grinstein:2008qi,Bernal:2009hd}.

  Both classes of mechanisms to enhance the strength of the electroweak 
phase transition require new physics below the TeV scale.  This is precisely
the energy regime that is currently being probed directly by 
high energy collider experiments and indirectly by lower-energy
 precision probes.  Some of the resulting experimental signals will 
be discussed in Section~\ref{sec:test}.


\section{Creating CP asymmetries and baryons
\label{sec:cpbg}}

  Baryons are created in EWBG through a combination of scattering, 
diffusion, and transfer reactions in the vicinity of the expanding 
bubble walls during the EWPT.  Sphaleron transitions are rapid in
the symmetric phase outside the bubbles, and they provide the requisite
source of baryon number violation.  However, sphalerons alone are not
enough; a net CP asymmetry is also needed to bias the sphalerons 
to to produce more baryons than antibaryons.

  The CP asymmetry relevant for EWBG is $n_L$, the excess number density 
of left-handed fermions over their antiparticles.  Such an asymmetry
can be created during the EWPT together with an equal opposite asymmetry 
in $n_R$ by CP-violating interactions in the bubble wall.  However, 
since sphalerons correspond to transitions between distinct $SU(2)_L$ vacua, 
only $n_L$ directly affects baryon creation.  The corresponding equation
for the baryon density $n_B$ is
\be
\label{eq:rhob1}
\partial_\mu j^\mu_B = -\frac{N_f}{2}
 \left[ k_\mathrm{ws}^{(1)}(T,x) n_B(x)
 +k_\mathrm{ws}^{(2)}(T,x) n_L(x)  \right]
\ee
where $j^\mu_B=(n_B,{\vec j}_B)$ is the baryon number current density,  
$x$ is the coordinate orthogonal to the bubble wall,
and $k_\mathrm{ws}^{(j)}(T,x)$ are the weak sphaleron rate constants
that account for the change in rate outside and inside the bubbles. 
Outside one has~\cite{Bodeker:1999gx}
\bea
\label{eq:ktout}
k_\mathrm{ws}^{(1)}(T,x)\Bigr\vert_\mathrm{out} &=& \mathcal{R}\times \frac{\Gamma_\mathrm{ws}}{VT^3} \\
\nonumber
k_\mathrm{ws}^{(2)}(T,x)\Bigr\vert_\mathrm{out} &=&  \frac{\Gamma_\mathrm{ws}}{VT^3}
\eea
with $\Gamma_\mathrm{ws}/V$ giving the weak sphaleron rate per unit volume 
for $N_f$ fermion families 
\bea
\frac{\Gamma_\mathrm{ws}}{VT^3}\Bigr\vert_\mathrm{out} & = & 6\kappa\alpha_W^5 T\\
\nonumber
\mathcal{R} & \simeq & \frac{15}{4}
\eea
and $\kappa\simeq 20$~\cite{Bodeker:1999gx}.
Deep inside the bubble these expressions become
\bea
\label{eq:ktin}
k_\mathrm{ws}^{(1)}(T,x)\Bigr\vert_\mathrm{in} & = & \frac{13 N_f}{2}\frac{\Gamma_\mathrm{ws}}{V T^3}\\
\nonumber
k_\mathrm{ws}^{(1)}(T,x)\Bigr\vert_\mathrm{in} & \approx & 0\\
\eea
The computation of $\Gamma_\mathrm{ws}/V$ inside the bubble has been discussed in Section \ref{sec:ewpt}. Note that Eq.~(\ref{eq:washoutb}) used in determining the washout factor is an approximation to Eq.~(\ref{eq:rhob1}) that neglects  ${\vec j}_B$ inside the bubble. 
It is clear from Eq.(\ref{eq:rhob1}) that without the chiral asymmetry
$n_L$, no net baryon charge will be created.


  The production of $n_L$ arises from interactions of particles with the bubble wall. The corresponding interaction rates are typically much faster than $k_\mathrm{ws}(T,x)$ so that in practice one may decouple Eq.(\ref{eq:rhob1}) from the system of equations that govern $n_L$ production. The latter encode a competition between several distinct dynamics: 
\begin{itemize}
\item[(a)] C- and CP-violating interactions with the bubble wall that lead to the generation of particle number asymmetries for one or more species 
\item[(b)] particle number-changing reactions that tend to drive the plasma toward chemical equilibrium; 
\item[(c)] flavor oscillations that result from off-diagonal mass-matrix elements
\item[(d)] scattering and creation-annihilation reactions that cause diffusion of the particle asymmetries ahead of the bubble wall and that push the system toward kinetic equilibrium in this asymptotic region. 
\end{itemize}

In the earliest work on this problem, these equations were reasonably assumed to have the form motivated by kinetic theory~\cite{Cohen:1994ss}
\be
\label{eq:nL1}
\partial_\mu j^\mu_k = -\sum_{a,\ell}\ \Gamma^{k\ell}_a \mu_\ell + S^\mathrm{CPV}_k\ \ \ ,
\ee
where $j^\mu_k= (n_k,\vec{j}_k)$ denotes the current density for a given species of particle \lq\lq $k$", $\Gamma_a^{k\ell}$ is the rate for a given reaction \lq\lq $a$ " that affects the chemical potentials of particle species \lq\lq $\ell$", and $S^\mathrm{CPV}_k$ is a CP-violating source term. 
  Diffusion arises in this formalism by making the diffusion {\em ansatz},
which relates the three-current to the gradient of the number density:
\beq
\vec{j}_k = D_k\,\vec{\nabla}n_k \ ,
\label{eq:diffeq}
\eeq
where $D_k(x,T)$ is the \emph{diffusion constant} for the particle species.  

  The reaction rate $\Gamma^{k\ell}_{a}$ terms in Eq.(\ref{eq:nL1}) 
couple different particle currents to one another.
For example, the top Yukawa interaction involving the third-generation 
SM quark doublet $Q$, the right-handed top quark $T$, 
and the Higgs doublet $H$ leads to a term of the form
\be
\label{eq:yuk1}
-\Gamma_Y \left(\mu_Q-\mu_T-\mu_H\right)
\ee
in the transport equation for $j^\mu_Q$ that couples to $j^\mu_T$ 
and $j^\mu_H$.  
CP-violation in this example is embodied by the source 
$S_Q^\mathrm{CPV}$, that tends to generate a non-zero number density $n_Q$.  
This number density would subsequently be diluted by the reaction
term of Eq. ~(\ref{eq:yuk1}) that transfers some of it to the 
$T$ and $H$ densities. 


Under the assumptions leading to Eq.(\ref{eq:nL1}), a robust determination of $n_L$ entails identifying all of the relevant particle species involved in the plasma dynamics, computing the relevant CP-violating sources and particle number changing reaction rates, and solving the resulting set of coupled equations.
Significant progress has been achieved over the years on all aspects of the problem. We first review studies of the particle number changing reactions.  Next, we discuss recent advances in deriving reliable CP-violating sources.  Throughout the discussion, we will use the MSSM as our primary example since this is the theory in which these issues have been studied in the greatest detail.

Before proceeding, we make a few preliminary comments regarding timescales. An important consideration in determining the relative importance of a given process in the production of $n_L$ is the associated time scale compared to the time scale for diffusion ahead of the bubble wall. As outlined in Refs.~\cite{Chung:2008aya,Chung:2009cb,Chung:2009qs}, in order for a given particle density to affect the dynamics of $n_L$ generation in the unbroken phase, that density must diffuse ahead of the advancing wall. The time it resides in the unbroken phase is the time it takes for the advancing wall to \lq\lq catch up" with the diffusing density. For a given time $t$, the diffusion length is $d_\mathrm{diff} =\sqrt{{\bar D}t}$, where ${\bar D}$ is an effective diffusion constant formed from an appropriate combination of individual diffusion constants $D_k$. The distance traversed by the wall is $d_\mathrm{wall}=v_w t$. Equating the two leads to the diffusion time scale: $\tau_\mathrm{diff}= {\bar D}/v_w^2$. Taking representative estimates for the effective diffusion constant ${\bar D}\simeq 50/T$ and wall velocity $v_w\simeq 0.05$ leads to $\tau_\mathrm{diff}\sim 10^4/T$. Any process $X$ having $\tau_X<< \tau_\mathrm{diff}$ must be included in the analysis of dynamics leading to $n_L$ production, whereas for $\tau_X >> \tau_\mathrm{diff}$ the process $X$ effectively decouples. In particular, the rate for electroweak sphaleron transitions implies $\tau_\mathrm{EW}>>\tau_\mathrm{diff}$ so that one may first solve the dynamics leading to $n_L$ and then use the latter as an input for the generation of $n_B$ via Eq.~(\ref{eq:rhob1}). 

Note that the wall velocity plays a significant role through its impact on $\tau_\mathrm{diff}$. As $v_w$ increases, other processes have less time to equilibrate before the relevant particle species are captured by the bubble. This effect is particularly important when $n_L$ is sourced by CPV interactions involving BSM degrees of freedom. In order for any charge asymmetry involving the latter to be transferred into a non-vanishing $n_L$ , the particle number-changing reactions that facilitate this transfer must take place more quickly than the time it takes for the wall to overtake the diffusion process.

\subsection{Particle number changing reactions}

  Tracking the full set of coupled particle number densities can be very
complicated.  For instance, in the MSSM with CP-violation in the Higgsino-gaugino sector, the resulting set of coupled equations can involve up to 30 or more components.  Fortunately, in many cases the scope of the problem can be significantly reduced.

  The most important simplification to the coupled equations is to relate algebraically the number densities of different particle species linked by reactions that are very fast compared to the other time scales of the problem (such as the diffusion time or the passage time of the bubble wall).  In our top Yukawa example in Eq.(\ref{eq:yuk1}), the corresponding reaction is typically very fast with $\tau_{Y_t}<<\tau_\mathrm{diff}$, implying that this process achieves chemical (as opposed to kinetic) equilibrium well before the $Q$, $H$, and $T$ densities are captured by the advancing wall. Consequently, it is an excellent approximation to set
\beq
\mu_Q - \mu_T-\mu_H = 0 \ .
\eeq
This reduces the number of coupled equations by one.

\begin{figure}[ttt]
\begin{center}
  \includegraphics[width=0.5\textwidth]{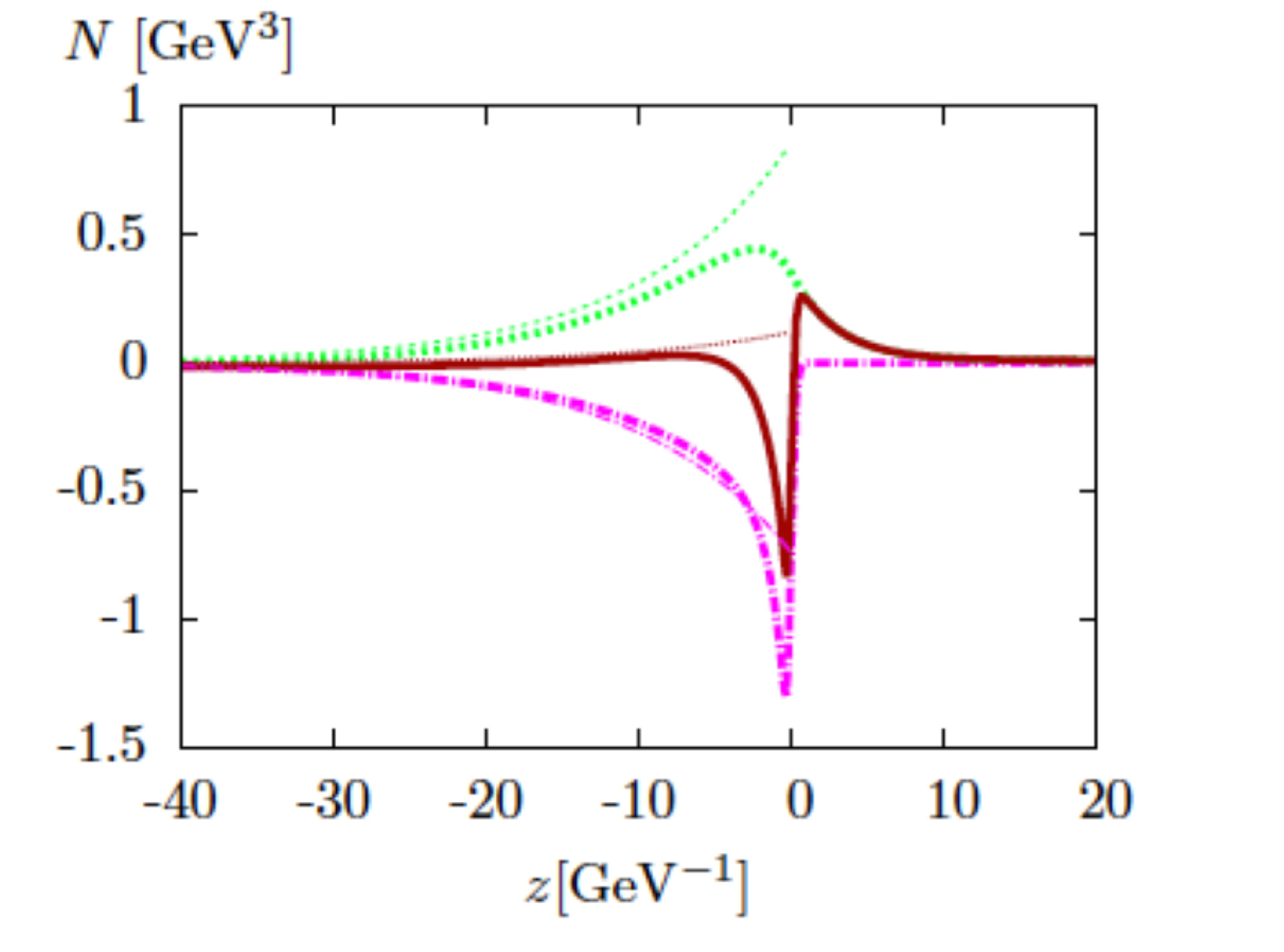}
\end{center}
\caption{\label{fig:nlprofile}
Particle number profiles with respect to the bubble wall ($z=0$) obtained in the MSSM~\cite{Chung:2009qs} . Pink, green, and red curves give, respectively, number densities of third generation left-handed quarks, third generation left handed leptons, and total $n_L$. Bubble wall interior corresponds to $z>0$. Thin curves represent results of an analytic approximation valid sufficiently far in front of the wall. Reprinted from D. J. H. Chung, B. Garbrecht, M. .J. Ramsey-Musolf and S. Tulin, \href{http://iopscience.iop.org/1126-6708/2009/12/067/}{JHEP 0912, 067 (2009)} with permission from JHEP.
}
\end{figure}

  A frequent simplification arises in supersymmetric theories such as the MSSM as a result of  \lq\lq superequilibration", wherein $\mu_P= \mu_{\tilde P}$ with $P$  and ${\tilde P}$ denoting a Standard Model particle  and its superpartner, respectively. Superequilibrium may arise either when a given supergauge interaction is fast compared to the timescale for diffusion ahead of the advancing bubble wall, or a chain of Yukawa reactions effectively yields $\mu_P= \mu_{\tilde P}$.  Exceptions may occur, however, when either the gauginos become heavy, thereby suppressing the corresponding supergauge reaction rates, or a chain of Yukawa reactions is kinematically blocked.

  Recently, the authors of Ref.~\cite{Chung:2009qs} carried out  an analysis of the particle number changing reactions for the MSSM under the assumptions leading to the general form of the transport equations in Eq.(\ref{eq:nL1}), obtaining numerical solutions to the full set of coupled transport equations.  Superequlibrium is found to obtain in significant regions of the parameter space.  Typical results for various particle number density profiles obtained in this work are given in Fig. \ref{fig:nlprofile}. Baryon number generation is governed by the size of the diffusion tail for $n_L$ (red curve) ahead of the advancing wall ($z<0$).

For scenarios involving more than a single Higgs doublet $H$, such as the MSSM, an additional situation involving Yukawa interactions may arise that can substantially alter the generation of $n_L$. In the Standard Model, the hierarchy of fermion Yukawa couplings implies that only $\tau_{Y_t}$ is much shorter than $\tau_\mathrm{diff}$ and that
one need consider only the $Q\!-\!T\!-\!H$ interactions indicated above. In a two Higgs doublet model (2HDM) scenario, however, the bottom quark (squark) and tau (stau) Yukawa couplings may be enhanced with respect to their SM values by $\tan\beta$, implying that the $\Gamma_Y$ involving these particles grows as $\tan^2\beta$. For $\tan\beta\gtrsim 5$ ($20$), the bottom (tau) Yukawa reactions  effectively equilibrate on a timescale shorter than the diffusion time scale, implying that one cannot neglect their effect on the coupled set of transport equations~\cite{Chung:2008aya}. 

\begin{figure}[ttt]
\begin{center}
  \includegraphics[width=0.5\textwidth]{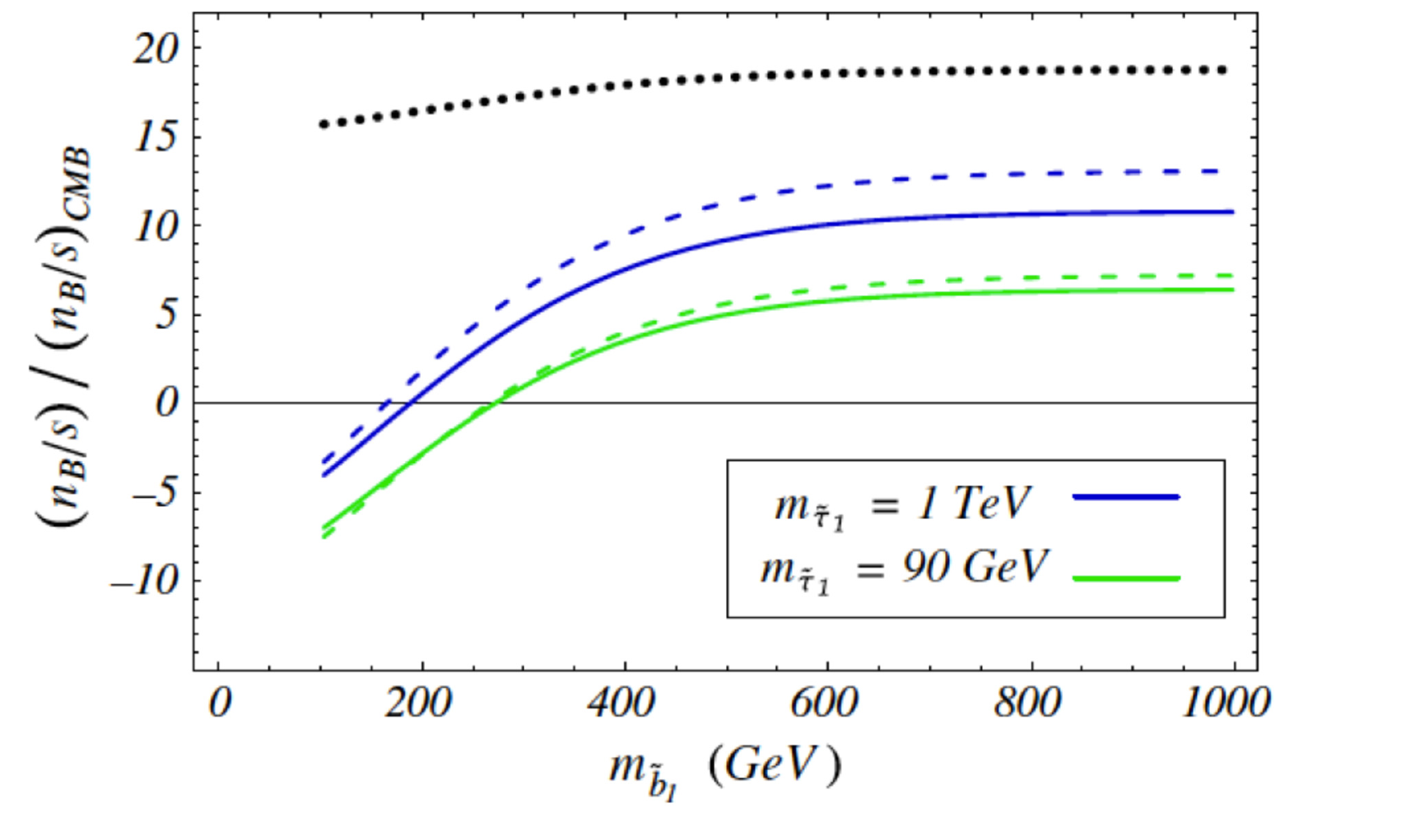}
\end{center}
\caption{\label{fig:yukawa}
Baryon asymmetry as a function of right-handed bottom squark mass in the MSSM, scaled to the value obtained from the cosmic microwave background~\cite{Chung:2008aya} The blue (green) curves correspond to right handed stau mass of 1 TeV (90 GeV), while black dotted line is obtained neglecting the bottom and tau Yukawa couplings. Solid curves give results of full numerical solution to transport equations, while  dashed line corresponds to an analytic approximation. A right-handed stop mass of 200 GeV has been assumed.
Reprinted with permission from D. J. H. Chung, B. Garbrecht, M. J. Ramsey-Musolf and S. Tulin, \href{http://link.aps.org/doi/10.1103/PhysRevLett.102.061301}{Phys. Rev. Lett. 102, 061301
(2009)}, copyright 2009 by the American Physical Society.
}
\end{figure}

  In the MSSM case, the net effect of the enhanced bottom Yukawa coupling can be to significantly suppress the final baryon asymmetry $n_B/s$ when the masses of the right-handed top and bottom squarks are nearly degenerate, or even lead to a sign change in the baryon asymmetry for $m_{\tilde b_R} <  m_{\tilde t_R}$. For sufficiently large $Y_\tau$ and light tau sleptons, a significant $n_L$ (and, thus, $n_B/s$) asymmetry can nevertheless be generated, as an initial Higgs-Higgsino asymmetry is transferred into the left-handed tau lepton sector through the tau Yukawa reactions~\cite{Chung:2008aya,Chung:2009cb}. These dynamics are illustrated in Fig.~\ref{fig:yukawa}, where one sees the vanishing of the baryon asymmetry for  degenerate right-handed tops (${\tilde t}_1$) and sbottoms (${\tilde b}_1$).

\subsection{CP-violating sources}

As the foregoing discussion makes clear, analyzing the detailed nature of the particle number changing interactions and their parameter dependence is essential for predicting the baryon asymmetry, even if the CP-violating sources have maximal strength. At the same time, deriving reliable CP-violating sources remains an on-going theoretical challenge that is part of a broader quest for a systematic treatment of the EWBG transport dynamics. 

   The earliest papers derived $S^\mathrm{CPV}$ using computations of quantum mechanical reflection and transmission from a barrier (the bubble wall). The idea was originally applied to the generation of a lepton number asymmetry during the EWPT by Cohen, Kaplan, and Nelson \cite{Cohen:1990py, Cohen:1990it} and subsequently applied directly to the baryon sector by Farrar and Shaposhnikov \cite{Farrar:1993sp,Farrar:1993hn,Farrar:1994kf} and others (see, {\em e.g.} Refs.~\cite{Joyce:1994bi,Joyce:1994zn}). The resulting baryon asymmetry within the Standard Model, computed using this approach, was generally considered to be too small to account for the observed value~\cite{Gavela:1993ts,Huet:1994jb}, so attention turned to scenarios involving physics beyond the Standard Model (BSM). A significant advance appeared in the incorporation of diffusion~\cite{Cohen:1994ss}, with the resulting framework remaining the state-of-the-art for the better part of a decade when applied to BSM models (see, \emph{e.g.}~\cite{Carena:1996wj,Carena:1997gx,Huet:1995sh,Cline:1995dg,Funakubo:1996dw,Davoudiasl:1997jh,Cline:2000nw}). 

\begin{figure}[ttt]
\begin{center}
  \includegraphics[width=0.5\textwidth]{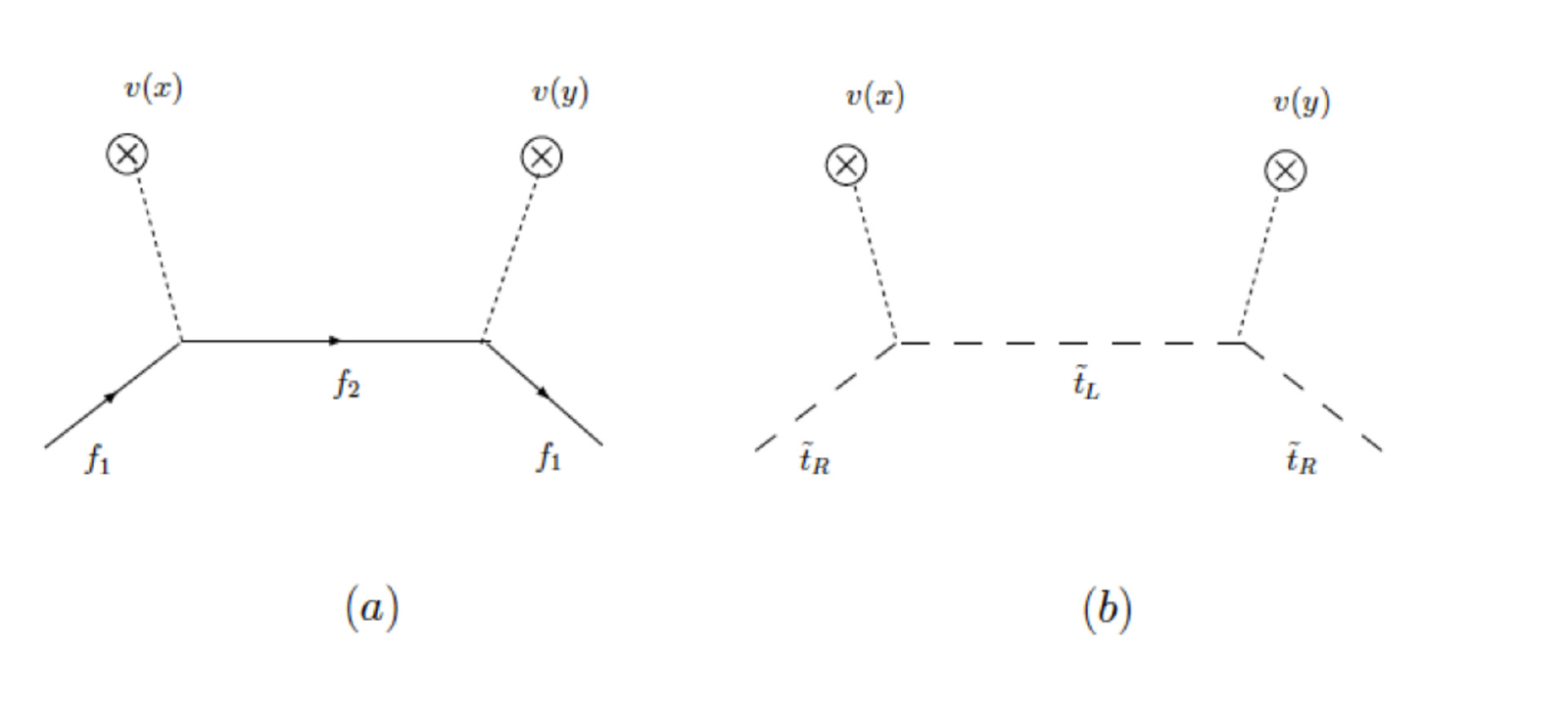}
\end{center}
\caption{\label{fig:vev}
Contributions to CP-violating sources in the VEV-insertion approximation\cite{Lee:2004we}. Here, $v(x)$ denotes appropriate combination of scalar background fields (VEVs) at spacetime position $x^\mu$ interacting with (a) fermions $f_{1,2}$ (b) scalar to quarks ${\tilde t}_{L,R}$.
Reprinted with permission from 
C. Lee, V. Cirigliano and M. J. Ramsey-Musolf, \href{http://link.aps.org/doi/10.1103/PhysRevD.71.075010}{Phys. Rev. D 71, 075010 (2005)},
copyright 2005 by the American Physical Society.
}
\end{figure}

  It was subsequently realized that the essentially Markovian nature of this framework omitted potentially important \lq\lq memory effects" that could lead to further enhancements of $n_L$~\cite{Riotto:1995hh,Riotto:1997vy,Riotto:1998zb}. Using the VEV insertion approximation depicted in Fig.~\ref{fig:vev}, where the Higgs VEV-dependent off-diagonal entries of the mass matrix are treated as perturbative interactions, Ref.~\cite{Riotto:1998zb} computed $S^\mathrm{CPV}$ in the MSSM for Higgsino and top squark transport near the bubble wall. As shown in the figure, the inputs for these sources involve two distinct particle species, either the Higgsinos and electroweak gauginos in the fermionic source case or the left- and right-handed top squarks in the scalar case. When the masses of the two species become nearly degenerate, $S^\mathrm{CPV}$ becomes resonantly enhanced. The presence of this feature is phenomenologically important, as an enhanced source will generate a particle asymmetry more effectively for a given CP-violating phase. Given the stringent limits on CP-violating phases implied by the non-observation of the electric dipole moments (EDMs) (see below), the resonant source enhancement is essential for viable EWBG in many instances.  The same resonant enhancement was also found to apply to the CP-conserving parts of Fig.~\ref{fig:vev}, which increases the relaxation rate of the non-vanishing densities sourced by the CP-violating parts, reducing 
the strength of the resonant enhancement of the baryon asymmetry 
but not completely mitigating it~\cite{Lee:2004we}.

\bigskip

A rigorous treatment of the spacetime evolution of particle densities that incorporates the memory effects is properly made using the Schwinger-Keldysh Closed Time Path (CTP) formulation of non-equilibrium quantum field theory\cite{Schwinger:1960qe,Keldysh:1964ud,Mahanthappa:1962ex,Bakshi:1962dv}  (for a pedagogical discussion in the context of EWBG, see Ref.~\cite{Lee:2004we}). In contrast to the \lq\lq in-out" matrix elements relevant to scattering processes, analysis of the spacetime evolution of densities in the plasma requires study of the \lq\lq in-in" matrix elements, appropriately averaged over a thermal ensemble. The corresponding evolution involves four Greens functions
\be
\label{eq:Gtilde}
{\tilde G}(x,y) = \left(
\begin{array}{cc}
G^t(x,y) & -G^<(x,y)\\
G^>(x,y) & -G^{\bar t}(x,y)
\end{array}\right)\ \ \ ,
\ee
where for a complex scalar field $\phi(x)$
\begin{eqnarray}
G^>(x,y) & = \langle \phi(x) \phi^\dag(y)\rangle_T \\
G^<(x,y) & = \langle \phi^\dag(y) \phi(x)\rangle_T 
\end{eqnarray}
while $G^t(x,y)=\theta(x_0-y_0) G^>(x,y) +\theta(y_0-x_0) G^<(x,y)$ and $G^{\bar t}(x,y)=\theta(x_0-y_0) G^<(x,y) +\theta(y_0-x_0) G^>(x,y)$.
Here, the subscript $T$ indicates an appropriate ensemble average. The fully interacting Greens functions ${\tilde G}(x,y)$ satisfy a pair of Schwinger-Dyson equations
\begin{eqnarray}
\label{eq:sd1}
{\tilde G}(x,y) &=& {\tilde G}(x,y)_0 +\int\, d^4z {\tilde G}(x,z)_0 {\tilde\Pi}(z,y) {\tilde G}(y,z)\\
\label{eq:sd2}
{\tilde G}(x,y) &=& {\tilde G}(x,y)_0 +\int\, d^4z {\tilde G}(x,z) {\tilde\Pi}(z,y) {\tilde G}(y,z)_0
\end{eqnarray}
where the \lq\lq 0" subscript indicates the free Greens function and ${\tilde\Pi}$ is a matrix of self-energy functions corresponding to Eq.(\ref{eq:Gtilde}). Acting with the free equation of motion operator ( {\em e.g.}, $\partial_x^2+m^2$ for a scalar field) on Eqs.~(\ref{eq:sd1},\ref{eq:sd2}), taking the difference of the two resulting equations, and considering the \lq\lq $<$" component in the limit $x\to y$ yields the transport equation
\begin{eqnarray}
\nonumber
\partial_\mu j^\mu(X) &=& \int d^3z\ \int_{-\infty}^{X_0}\ \Bigl[\Pi^>(X,z) G^<(z,X)-G^>(X,z) \Pi^<(z,X)\\
&&+G^<(X,z) \Pi^>(z,X) -\Pi^<(X,z) G^>(z,X)\Bigr]\ \ \ ,
\end{eqnarray}
where the effect of all interactions are contained in the convolution of the Greens functions and self-energies on the right hand side. In the VEV insertion approximation, CP-violating and CP-conserving interactions of the scalar field $\phi$ with  spacetime-varying Higgs background field (the bubble wall) are contained the self-energies. Particle number changing interactions, as well as those associated with scattering and creation/annihilation, also live in the ${\tilde\Pi}$. Note that the source for the divergence of the particle number current contains an integral over the history of the system, leading to the presence of memory effects as emphasized in Ref.~\cite{Riotto:1998zb}.

Subsequent work has attempted to refine this formulation,
particularly within the MSSM.  The authors of Refs.~\cite{Carena:2000id,Carena:2002ss} endeavored to go beyond the VEV insertion approximation by noting that, when resummed to all orders in the VEVs,  the interactions of Fig.~\ref{fig:vev} lead to spacetime-dependent mass matrices for the supersymmetric particles. In the case of stops, for example, the corresponding term in the Lagrangian is
\be
\label{eq:stopmass}
\mathcal{L}_{\tilde t}^\mathrm{mass} = 
\left( 
\begin{array}{cc}
{\tilde t}_L^\ast & {\tilde t}_R^\ast
\end{array}
\right)
\left(
\begin{array}{cc}
M_{LL}^2(x) & M_{LR}^2(x) \\
M_{LR}^{2}(x)^\ast & M_{RR}^2(x) 
\end{array}
\right)
\left(
\begin{array}{c}
{\tilde t}_L \\
{\tilde t}_R
\end{array}
\right)\ \ \ ,
\ee
where
\begin{eqnarray}
M_{LL}(x) & = & M_Q^2+ m_t^2(x) + \Delta_t(x)\\
M_{RR}(x) & = & M_T^2+ m_t^2(x) + \Delta_t(x)\\
M_{LR}^2(x) & = & y_t v(x) \left[A_t \sin\beta(x) - \mu \cos\beta(x)\right]\ \ \ .
\end{eqnarray}
Here, $M_Q$ and $M_T$ are, respectively, the spacetime-independent third generation left- and right-handed stop mass parameters; $m_t(x)=y_t v(x)/\sqrt{2}$ is the spacetime dependent top mass, with top Yukawa coupling $y_t$ and spacetime-dependent VEV $v(x) =\sqrt{v_u^2(x) + v_d^2(x)}$; $\Delta_t(x)$ a function of $\beta(x) = \tan^{-1}[v_u(x)/v_d(x)]$, $v(x)$, and the weak mixing angle; and where 
$\mu$ and $y_t A_t$ are supersymmetric Higgs mass and stop triscalar couplings, respectively. Note that in general $\mathrm{Arg}(\mu A_t)\not=0$, so that the off-diagonal entry $M_{LR}^2(x)$ effectively contains a spacetime-dependent CP-violating phase. 
At each point in spacetime, then, the propagating degrees of freedom -- the mass eigenstates ${\tilde t}_j$ ($j=1,2$) -- are related to the weak interaction eigenstates ${\tilde t}_a$ ($a=L,R$) by a unitary transformation $U(x)$
\be
\label{eq:stoptransf}
{\tilde t}_j = U(x)_{ja}^\dag {\tilde t}_a\ \ \ .
\ee

  The authors of Refs.~\cite{Carena:2000id,Carena:2002ss} computed the current densities for the ${\tilde t}_a$ ($a=L,R$) by expanding the mass-squared matrix $M^2_{ab}(x)$ in Eq.(\ref{eq:stopmass}) about a given point $z^\mu$ by a radius $r$, diagonalizing $M^2_{ab}(z)$ using Eq.(\ref{eq:stoptransf}), solving for the locally-free Greens functions ${\tilde G}^0(z-r/2,z+r/2)$, and treating the first order term in the expansion of the mass-squared matrix,
\be
\label{eq:firstorderx}
r^\lambda \frac{\partial M^2}{\partial x^\lambda} \ ,
\ee
as an interaction in the CTP Schwinger-Dyson equations, Eqs.(\ref{eq:sd1}~\ref{eq:sd2}). The resulting solution for the LH stop current $j^\lambda_{\tilde t_L}$ then drives the creation of non-vanishing baryon number via the EW sphalerons, as per Eqs.~(\ref{eq:nL1},\ref{eq:rhob1}). Similar methods were applied  to determine the chargino current $j^\lambda_{\tilde \chi^\pm}$.

The effect of including the fully re-summed VEVs and the first order interaction of Eq.(\ref{eq:firstorderx}) is to reduce the strength of the resonance. As an illustration for the chargino sources with $M_2=200$ GeV, $M_A=150$ GeV, and $\mathrm{Arg}(\mu A_t)=\pi/2$, one finds that for $|\mu|\approx M_2$ the baryon asymmetry $Y_B$ is enhanced by about a factor of seven relative to its off-resonance magnitude at $|\mu|=100$ GeV. This result represents about a factor of two reduction in $Y_B$ compared to the result obtained using the VEV insertion approximation in Ref.~\cite{Carena:1996wj}, and is considerably smaller still than obtained in Refs.~\cite{Riotto:1998zb,Lee:2004we} . Nonetheless, it suggests that obtaining the observed value of $Y_B$ in the MSSM {\em via} chargino and neutralino sources would be viable even when a VEV resummation is performed.

\begin{figure}[ttt]
\begin{center}
  \includegraphics[width=0.5\textwidth]{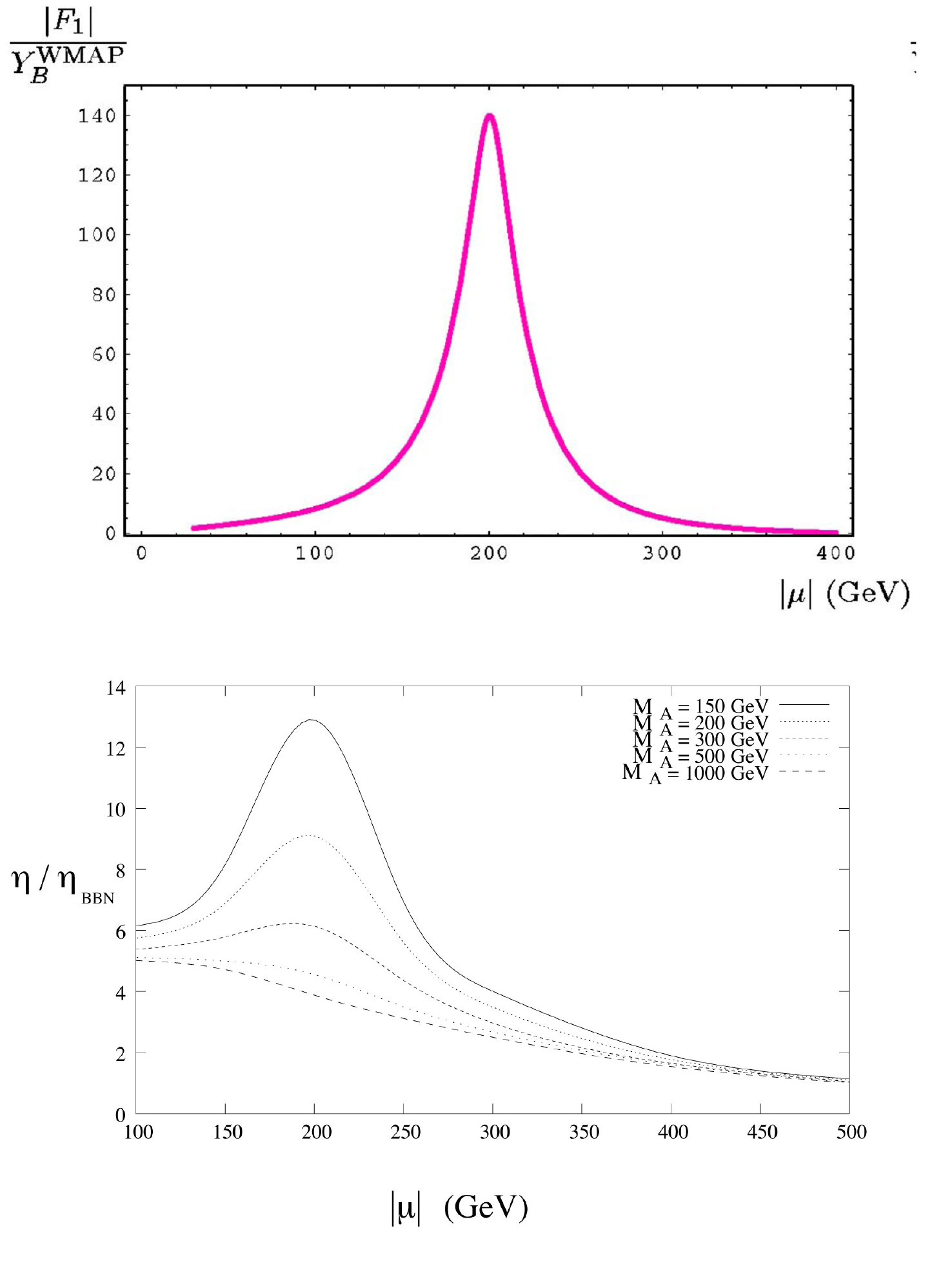}
\end{center}
\caption{\label{fig:resonance}
Baryon asymmetry in the MSSM. Top panel:  computed using VEV-insertion approximation and relaxation terms while including contributions from both charged and neutral Higgsinos and electroweak gauginos\cite{Lee:2004we}. Bottom panel: including only charged Higgsino and wino contributions but following the VEV resummation approach of Refs.~\cite{Carena:2000id,Carena:2002ss}, from Ref~\cite{Balazs:2004ae}.
Top panel reprinted with permission from C. Lee, V. Cirigliano and M. J. Ramsey-Musolf, \href{http://link.aps.org/doi/10.1103/PhysRevD.71.075010}{Phys. Rev. D 71, 075010 (2005)}, copyright 2005 by the American Physical Society.  Bottom panel reprinted with permission from C. Balazs, M. S. Carena, A. Menon, D. E. Morrissey and C. E. M. Wagner, \href{http://link.aps.org/doi/10.1103/PhysRevD.71.075002}{Phys. Rev. D 71,
075002 (2005)}, copyright 2005 by the American Physical Society.
}
\end{figure}

Parallel efforts on the resummation problem were carried out in Refs.~\cite{Kainulainen:2001cn,Prokopec:2003pj,Prokopec:2004ic}, adopting the framework of the Kadanoff-Baym equations~\cite{KadBay62} that are deduced from the Schwinger-Dyson equations. (For a related effort using the Bogoliubov approach, see Ref.~\cite{Garbrecht:2002pd}). Applying this analysis to the MSSM and performing a 2-flavor analysis  with 
approximate treatment of the off-diagonal densities the authors of  Refs.~\cite{Konstandin:2004gy,Konstandin:2005cd} concluded that a proper treatment of flavor oscillations as well as a VEV resummation vastly suppresses $Y_B$ in the MSSM, rendering it unviable in light of current EDM limits on the CP-violating phases.  
 
  To explain the arguments behind these conclusions, the subsequent response by the authors of Refs.~\cite{Cirigliano:2009yt,Cirigliano:2011di}, and the parallel work by the authors of Refs.~\cite{Herranen:2008hi,Herranen:2008hu,Herranen:2008di,Herranen:2010mh}, we present here a specific example involving scalar particles, which avoids the complications associated with spin.   
Our example consists of a two-flavor scalar field theory.
In the presence of a space-time dependent mass-squared matrix, the performing the flavor rotation of Eq.(\ref{eq:stoptransf}) leads to the Lagrangian in terms of local mass eigenstates $\Phi=(\phi_1,\phi_2)$
\begin{eqnarray}
\label{eq:Lscalar1}
\mathcal{L}_\mathrm{scalar} = {\partial_\mu\Phi}^\dag \partial^\mu\Phi - \Phi^\dag {\hat M}^2(x) \Phi -\Phi^\dag\Sigma^\mu\partial_\mu \Phi\\
\nonumber
 + (\partial_\mu \Phi^\dag) \Sigma^\mu\ \Phi
+\Phi^\dag \Sigma^\mu \Sigma_\mu \Phi+ \mathcal{L}_\mathrm{int}\ \ \ ,
\end{eqnarray}
where ${\hat M}^2(x)=\mathrm{diag}[ m_1^2(x), m_2^2(x)]$, 
\be
\Sigma^\mu = U^\dag(x) \partial^\mu U(x) \ ,
\ee
and $\mathcal{L}_\mathrm{int}$ contains interactions of $\Phi$ with other particles in the plasma that lead to chemical and kinetic equilibrium as well as diffusion ahead of the bubble wall. 

  The effects of CP-violation are encoded in $\Sigma^\mu$ that contains, in particular, a spacetime derivative of the effective CP-violating phase associated with the off-diagonal elements of the un-diagonalized matrix $M^2(x)_{ab}$. Applying the equation of motion associated with all terms in Eq.(\ref{eq:Lscalar1}) except  $\mathcal{L}_\mathrm{int}$ to the Schwinger-Dyson equations for the ${\tilde G}$ leads to
\begin{eqnarray}
\nonumber
\left[\partial_x^2+{\hat M}^2(x) + 2 \Sigma_\mu\partial^\mu_x +\Sigma^2 + (\partial\cdot\Sigma)\right]G^{<}(x,y)\\
\label{eq:sd2x}
=-i\int\ d^4 z \left[{\tilde\Pi}(x,z){\tilde G}(z,y)\right]^{<}\ \ \ ,
\end{eqnarray}
with a similar equation holding for the operators acting on the $y$-coordinate. Taking the sum and difference of these two equations yields the constraint and kinetic equations, respectively. The former determines the dispersion relations for the propagating degrees of freedom, while the latter governs their dynamics. Extensive reviews of the related formalism can be found in Refs.~\cite{Calzetta:1986cq,Berges:2004yj}. 

For the sake the present discussion, we concentrate only on the kinetic equation. In doing so, it is convenient to consider the kinetic equation for the Wigner transform of the $G^<(x,y)$:
\be
G^<(k,X)= \int\ d^4r e^{ik\cdot r}\ G^<(x,y)
\ee
with $r=x-y$ and $X=(x+y)/2$. The kinetic equation is
\begin{eqnarray}
\nonumber
2 k\cdot\partial_X G^<(k,X) &=& e^{-i\diamond}\Bigl\{-i\left[{\hat M}^2(X)-2ik\cdot\Sigma(X)+\Sigma^2(X), G^<(k,X)\right]\\
\label{eq:kinetic}
&&+\mathcal{C}(k,X)\Bigr\}\ \ \ ,
\end{eqnarray}
where $\mathcal{C}(k,X)$ denotes the so-called \lq\lq collision term" involving products of the $G$ and $\Pi$ Wigner transforms and the \lq\lq diamond operator" is defined by
\be
\diamond(AB)=\frac{1}{2} \left(  \frac{\partial A}{X^\mu}\frac{\partial B}{k_\mu}-\frac{\partial B}{X^\mu}\frac{\partial A}{k_\mu} \right)\ \ \ .
\ee
Note that Eq.(\ref{eq:kinetic}) and its partner constraint equation represent a set of coupled integral-differential equations in the space of Greens functions for the local mass eigenstates. For simplicity we have suppressed the corresponding indices that would appear, for example, in the term
\begin{eqnarray}
\label{eq:comm}
\nonumber
\left[-2ik\cdot\Sigma(X), G^<(k,X)\right]_{ij} &=& -2ik_\mu\Bigl[\Sigma^\mu(X)_{i\ell}G^<(k,X)_{\ell j}\\
&& -G^<(k,X)_{i\ell }\Sigma^\mu(X)_{\ell j}\Bigr]\ \ \ .
\end{eqnarray}

The different terms in Eq.(\ref{eq:kinetic}) embody various key physical dynamics that govern the creation of a net particle density ahead of the bubble wall: (a) the $[{\hat M}^2,G^<]$ term gives rise to flavor oscillations, much in analogy with those observed or neutrino oscillations; (b) the commutators involving the field $\Sigma^\mu$ contain the effects of the spacetime-dependent CP-violating phases that are essential for the generation of a non-vanishing number density; (c) the collision term $\mathcal{C}$ includes the effects of scattering and particle creation/annihilation that gives rise to diffusion ahead of the bubble wall and thermalization within the plasma, thermal corrections to the masses and widths of the propagating degrees of freedom, and the particle number changing reactions discussed above. Note that Eq.(\ref{eq:kinetic}) is exact to all orders in the spacetime-varying Higgs VEVs, is implied by the $X^\mu$-dependence of the mass-squared matrix, the field $\Sigma^\mu$, and $\mathcal{C}$. The  $X^\mu$-dependence of ${\hat M}^2(X)$ also gives rise to the \lq\lq semi-classical force" term that was first derived in this context for fermionic systems in Ref.~\cite{Kainulainen:2001cn}  and that generates a CP-violating source term even for a single flavor situation. 

Obtaining exact solutions of the kinetic and constraint equations is a daunting task.  Recent progress has been achieved in obtaining approximate solutions by expanding the various terms in powers of  different scale ratios $\epsilon_j$ appropriate to the plasma dynamics (see Refs.~\cite{Lee:2004we,Cirigliano:2009yt,Cirigliano:2011di} for a discussion). The scales include (a) the typical quasiparticle frequency $\omega_\mathrm{int}$ or deBroglie wavelength $L_\mathrm{int}$ that is intrinsic to its free propagation. Both scales are set by the plasma temperature (assuming local thermalization): $\omega_\mathrm{int}\sim L_\mathrm{int}\sim T$. (b) The wall thickness $L_w$ or the timescale $\tau_w$ over which the particle masses vary appreciably at a local point in space as the bubble expands; (c) The typical frequency of flavor oscillations, $\omega_\mathrm{osc}$ given by the difference of the local eigenfrequencies $|\omega_1-\omega_2|$  or the associated time and length scales, $\tau_\mathrm{osc}\sim L_\mathrm{osc}/c\sim  \omega_\mathrm{osc}^{-1}$; the length or time scales associated with the collisional interactions, $\tau_\mathrm{col}\sim L_\mathrm{col}/c$. The latter set the rates for diffusion, kinetic and chemical equilibration, and thermal damping (widths) of the quasiparticles; and (d)the chemical potentials associated with various particle species $\mu_j$. Typically one finds the following ratios of scales to be significantly less than one:
\be
\epsilon_w=\frac{L_\mathrm{int}}{L_w}\qquad \epsilon_\mathrm{col}= \frac{L_\mathrm{int}}{L_\mathrm{col}}\qquad\epsilon_\mu=\frac{\mu}{T}\ \ \ .
\ee

The parameter $\epsilon_w$ is particularly important for electroweak baryogenesis, as it must be non-zero in order for the coupled set of kinetic equations to generate a non-vanishing particle density. Since typically one finds $L_w\sim\mathcal{O}(20/T)$, implying that  $\epsilon_w\lesssim 0.5$. 
In this regime, an expansion in powers of $\epsilon_w$ -- sometimes referred to as the \lq\lq gradient expansion" -- is justified, and most recent work has focused on solutions obtained to first order in this small parameter\footnote{This validity of the gradient or $\epsilon_w$ expansion corresponds to the \lq\lq thick wall" regime in earlier work.} 
 In general, one also finds that $\epsilon_\mathrm{col}<<1$ and $\epsilon_\mu << 1$, leading to well-defined approximations by expanding in these quantities. In contrast, the ratio of \lq\lq intrinsic" and oscillation length scales, $\epsilon_\mathrm{osc}=L_\mathrm{int}/L_\mathrm{osc}$ depends strongly on the input parameters of the underlying Lagrangian. In the case of resonant baryogenesis, $\omega_\mathrm{osc}$ is relatively small, and one finds that the resonant enhancement occurs when the corresponding length scale $L_\mathrm{osc}\sim L_w >> L_\mathrm{int}$. In this case, one may also expand in $\epsilon_\mathrm{osc}$ (though some terms must be resummed to all orders in $\omega_\mathrm{osc}$ in order to maintain consistency with the continuity equation). 

The work of Refs.~\cite{Konstandin:2004gy,Konstandin:2005cd} yielded the first solutions of the kinetic equations for the MSSM under these approximations that take into account interplay of both the fully resummed spacetime varying VEVs with the off-diagonal densities. Applying the power counting in $\epsilon_w$ indicated above, the argued that the off-diagonal contributions to the RHS of Eq.(\ref{eq:comm}) for the diagonal densities $G^<(k,X)_{ii}$ arise beyond leading non-trivial order, allowing one to neglect their effect as CPV sources for the diagonal terms. The corresponding numerical results then indicated that the value of $Y_B$ in the resonant regime is substantially smaller than obtained in earlier work. Given the limits on CP-violating phases from EDM searches (see below), one would then conclude that EWB in the MSSM cannot yield the observed baryon asymmetry, contrary to the implications of earlier work. 

Recently, however, the authors of Ref.~\cite{Cirigliano:2011di} pointed out, using a simpler schematic two-flavor scalar field theory, that the approximations used in Refs.~\cite{Konstandin:2004gy,Konstandin:2005cd} do not consistently implement the power counting in the $\epsilon_w$. In particular, the solution to Eqs.~(\ref{eq:kinetic}) involve a integral of the terms in Eq.(\ref{eq:comm}) over a distance scale of order $1/L_w$. Although the fields $\Sigma^\mu$ are nominally $\mathcal{O}(\epsilon_w)$, this integral compensates for this nominal $1/L_w$ suppression. Thus, all contributions from the commutator in Eq.(\ref{eq:comm}), including those involving off-diagonal terms $k\cdot\Sigma_{12}G^<_{21} -G^<_{12}k\cdot\Sigma_{21}$ that contribute to the evolution of the diagonal Greens functions $G^<_{11}$ must be kept to leading non-trivial order in the $\epsilon_j$. Dropping these off-diagonal contributions, as in Refs.~\cite{Konstandin:2004gy,Konstandin:2005cd} , removes the dominant CP-violating source for the diagonal Greens functions that characterize the diffusion of particle number ahead of the advancing wall and leads to the suppressed asymmetry as obtained in that work. On the other hand, the consistent solution that retains these terms displays the substantial resonant enhancement that was noted in earlier works obtained with the VEV-insertion approximation. 

It is important to emphasize that the most recent work on this problem has been completed using schematic two-flavor scalar field models, and not the MSSM for which CP-violating dynamics involving fermion fields (Higgsinos and gauginos) is the most viable mechanism for successful baryogenesis in this scenario. In addition, the authors of Ref.~\cite{Herranen:2010mh} -- building on earlier work~\cite{Herranen:2008hi,Herranen:2008hu,Herranen:2008di} and concentrating on the time-dependent, spatially homogeneous background field problem -- have argued that contributions from so-called \lq\lq fast modes" or \lq\lq coherence shells"  may have a significant impact on the magnitude of the particle asymmetries. In contrast to solutions of the constraint equations that satisfy particle-like dispersion relations (quasiparticle modes), the coherence shells correspond to regions of vanishing space or time components of the momentum and generate singular contributions to the spectral densities. A consistent treatment of these modes, which can mix with the quasiparticle modes, appears to require a resummation to all orders in $\epsilon_w$. The implications for the full spacetime-dependent problem remain to be analyzed. 
On the other hand, the progress achieved in Refs.~\cite{Cirigliano:2009yt,Cirigliano:2011di} represents a significant advance, not only for the implementation of a consistent power counting, but also for the full inclusion of the collision term $\mathcal{C}(k,X)$ that governs the evolution of kinetic equilibrium ahead of the advancing wall. Previous work had largely implemented equilibrium as an {\em ansatz}, rather than obtaining it as a direct solution of the dynamics. 

Significant formal and phenomenological work remains to be completed. A  full application to the dynamics of fermions -- including resolution of questions involving the\lq\lq semi-classical force" term, the coherence shells, and off-diagonal densities -- as well as a more realistic set of thermalizing interactions in the collision term are among the open theoretical problems. Phenomenological application to realistic scenarios like the MSSM that involve a more extensive set of coupled integral-differential equations, including both the network of particle number changing reactions as in Refs.~\cite{Chung:2009qs,Chung:2008aya,Cirigliano:2006wh} as well as the consistent treatment of flavor oscillations and spacetime varying background fields in the presence of the thermalizing interactions in the plasma, will entail significant effort. Nonetheless, given the recent theoretical progress and indications that resonant EWB remains a viable mechanism, this investment of effort is likely to have a significant impact on the field.

\section{Testing electroweak baryogenesis
\label{sec:test}}

  Electroweak baryogenesis requires new particles and interactions
to obtain a strongly first-order electroweak phase transition
and to provide sufficient CP violation.  These new particles cannot be
much heavier than the electroweak scale, and they must couple
significantly to the Higgs field.  Together, these properties
imply that such new particles will lead to observable effects
in upcoming high-energy and high-precision experiments.

The prospect for observing new particles directly at the LHC and indirectly through
high-sensitivity, low-energy studies of CP-violating observables imply that EWBG is a generally
testable and, therefore, falsifiable, baryogenesis scenario. In this respect, it contrasts with other scenarios that
typically involve higher scales, such as standard thermal leptogenesis or Affleck-Dine baryogenesis. In what follows, we summarize some
of the primary experimental tests of EWBG, focusing largely but not exclusively on the MSSM as an illustration. We consider tests that may be 
performed each of the three-frontiers in particle physics: the high-energy frontier; 
the intensity frontier; and the cosmological 
frontier.

\subsection{The Intensity Frontier: CP-Violation and EWBG}
\label{sec:cpv}

In general, the most powerful probes of BSM CP-violation that are relevant for EWBG are searches for the permanent electric dipole moments (EDMs) of 
the electron, neutron, and neutral atoms. In all cases, only null results have been attained to date, implying stringent constraints on new sources of CP-violation. Limits also exist on the muon EDM, though it is considerably less constraining than the electron EDM limit, even after accounting for the $m_e/m_\mu$ suppression of the former with respect to the latter. At the most basic level, one expects the one-loop EDM of an elementary fermion $f$ generated by new field(s) of mass $M$ to go as
\be
\label{eq:edm0}
d_f\sim e \left(\frac{m_f}{M^2}\right)\ \frac{\alpha_k}{4\pi} \sin\phi\ \ \ 
\ee
where $\alpha_k$ is either the fine structure constant or strong coupling (evaluated at the scale $M$) and $\phi$ is a CP-violating phase associated with the new interactions. For $\alpha_k=\alpha_\mathrm{em}$ Eq.(\ref{eq:edm0}) gives
\be
\label{eq:edm2}
d_f\sim\sin\phi\ \left( \frac{m_f}{\mathrm{MeV}}\right)\ \left( \frac{1\ \mathrm{TeV}}{M}\right)^2 \times 10^{-26}\ \ecm\ \ \ .
\ee
The present limit on the EDM of the electron, $|d_e| < 10.5\times 10^{-28}$ $\ecm$~\cite{Hudson:2011zz} obtained from an experiment on the Yb-F molecule, then implies that 
\be
\label{eq:edm1}
|\sin\phi| \lesssim \left( M / 2\ \mathrm{TeV}\right)^2\ \ \ .
\ee
Similar constraints follow from the limits on the neutron~\cite{Baker:2006ts} and $^{199}$Hg atomic~\cite{Griffith:2009zz} EDMs:
\begin{eqnarray}
| d_n| & < & 2.9 \times 10^{-26}\quad \ecm\\
| d_A(^{199}\mathrm{Hg})| & < & 3.1 \times 10^{-29}\quad \ecm\ \ \ ,
\end{eqnarray}
assuming that any contributions from the QCD $\theta$-term interaction are sufficiently small that no cancellations between this source of SM CP-violation and that arising from new interactions occurs. Contributions from CP-violation associated with the SM CKM matrix first arise at three (four) loop order for $d_n$ and $d_A$ ($d_e$), implying effects at well-below the $10^{-30}\ecm$ level. The next generation of lepton, neutron, and neutral atom EDM searches aim to improve the level of sensitivity by up to two orders of magnitude in the short term, while efforts to reach even greater sensitivity with storage ring hadronic EDM searches are underway (for a recent summary of present plans, see {\em e.g.}, Ref.~\cite{ifwp}).

The  constraints implied in Eq.(\ref{eq:edm1}) generically render EWBG unviable. For the new particles to be sufficiently abundant in the electroweak plasma at $T\sim 100$ GeV, their masses should be lighter than $\sim 500$ GeV, implying $|\sin\phi|\lesssim 0.01$. In this case, the CPV sources in the transport equations discussed in Section \ref{sec:cpbg} are suppressed and EWBG becomes untenable as a result (cf Figs. \ref{fig:yukawa},\ref{fig:resonance}). There exist, however, several paths to evading the one-loop EDM constraints. In the MSSM, the one-loop EDMs contain one scalar ({\em e.g.} squark or slepton) and one fermionic superpartner (gaugino or Higgsino). By making one or the other species sufficiently heavy, the one-loop EDMs can be evaded, thereby relaxing the constraints of Eq.(\ref{eq:edm1}) on the CPV phase. At the same time, the other superpartner species may remain relatively light, enabling its interactions to generate the CPV sources for electroweak transport dynamics. Large differences in the scalar and fermion mass spectra have been motivated on other grounds recently, as in the case of \lq\lq split SUSY" models~\cite{Giudice:2004tc,Giudice:2005rz} that contain heavy first and second generation sfermions but relatively light Higgsinos and electroweak gauginos. The present generic LHC lower bounds on the masses of the gluinos and first and second generation squarks is at least consistent with this scenario. 

\begin{figure}[ttt]
\begin{center}
  \includegraphics[width=0.5\textwidth]{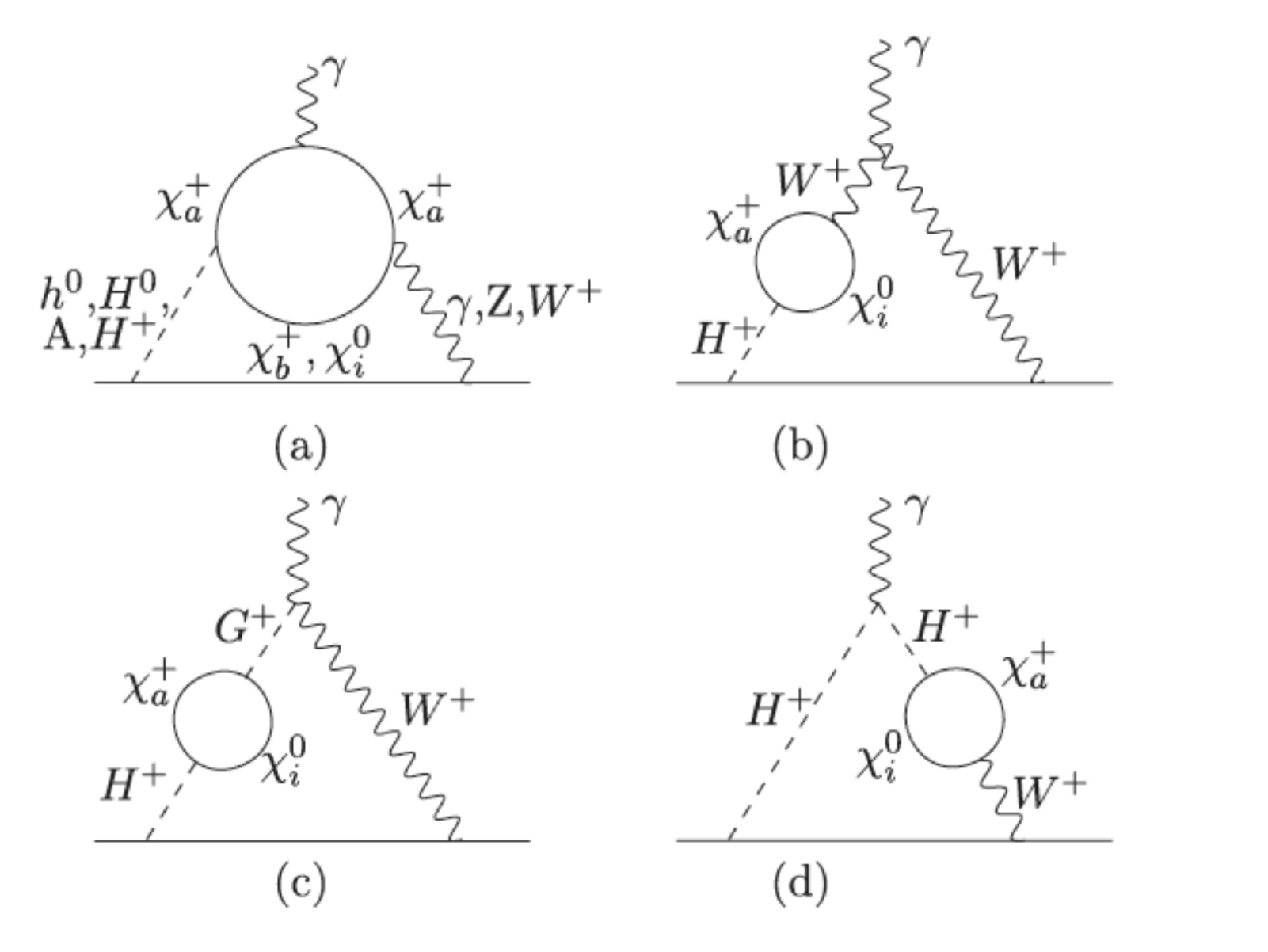}
\end{center}
\caption{\label{fig:edm}
Two loop EDMs in the MSSM (see Ref.~\cite{Li:2008kz} and references therein).
Reprinted with permission from Y. Li, S. Profumo and M. Ramsey-Musolf, \href{http://link.aps.org/doi/10.1103/PhysRevD.78.075009}{Phys. Rev. D 78, 075009 (2008)},
copyright 2008 by the American Physical Society.
}
\end{figure}

\begin{figure}[ttt]
\begin{center}
  \includegraphics[width=0.8\textwidth]{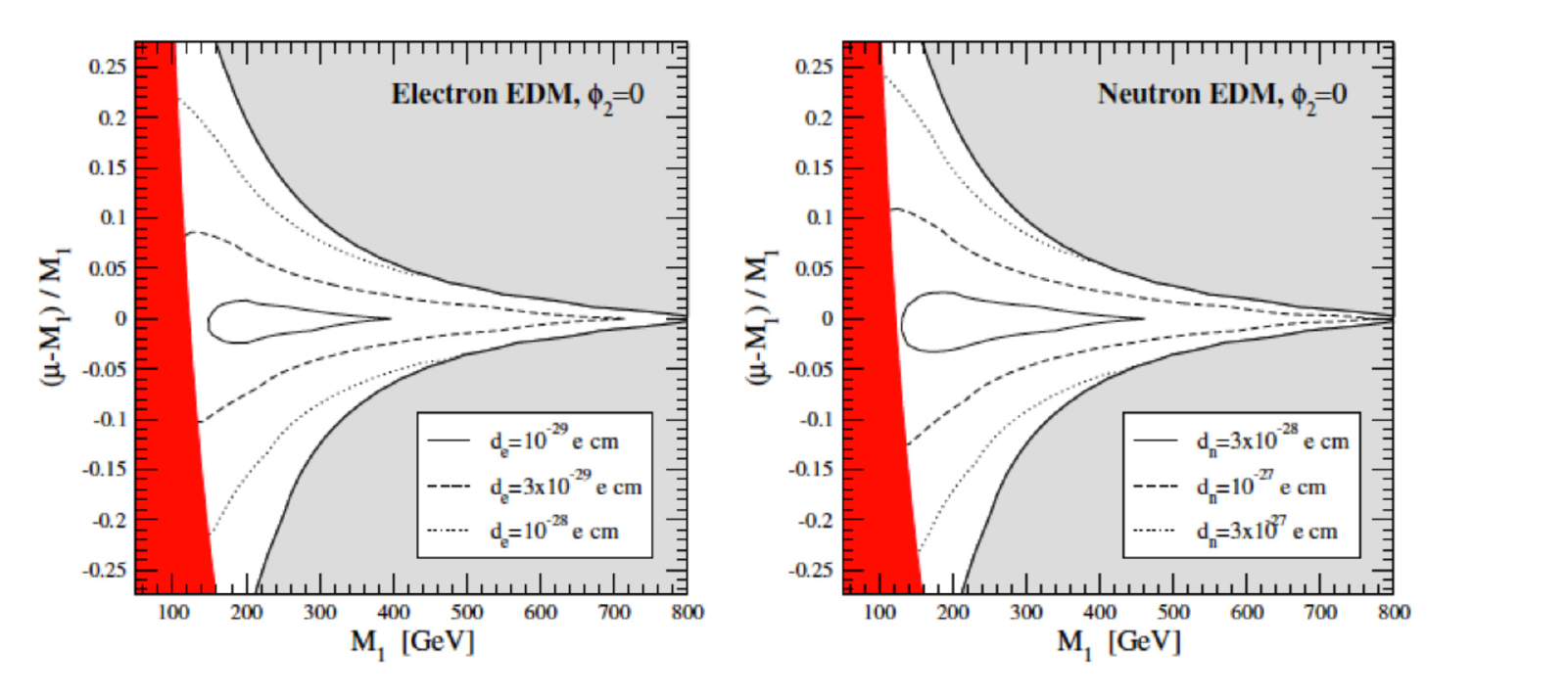}
\end{center}
\caption{\label{fig:mssmcpv}
Curves of constant $d_e$ and $d_n$ for MSSM EWBG in the limit of heavy sfermions~\cite{Cirigliano:2009yd}. For each curve, $\phi_1$ is set to the value giving the correct baryon asymmetry.  Reprinted from V. Cirigliano, Y. Li, S. Profumo and M. J. Ramsey-Musolf, \href{http://www.springerlink.com/content/g071711q4k130772/?MUD=MP}{JHEP 1001, 002 (2010)} with permission from JHEP.}
\end{figure}

From the standpoint of CPV for EWBG, one must still consider EDMs generated at two-loop level, as in the \lq\lq Barr-Zee" graphs of Fig.~\ref{fig:edm}. Recently, the full set of such diagrams were computed by the authors of Ref.~\cite{Li:2008kz} and the corresponding implications for MSSM baryogenesis delineated in Refs.~\cite{Li:2008ez,Cirigliano:2009yd}. Even with the two-loop suppression and the most optimistic CPV sources computed using the VEV-insertion approximation, only one CPV phase remains sufficiently unconstrained as to remain a potentially viable driver of MSSM EWBG: the relative phase of the bino soft mass parameter $M_1$, the soft parameter $b$ and the supersymmetric Higgs mass parameter $\mu$: $\phi_1=\mathrm{Arg}(M_1\mu b^\ast)$.   Since $\phi_1$ is associated with the presence of the bino degrees of freedom, it only enters the two-loop graphs involving the $\chi^0_k$ and the exchange of a $(W^\pm, H^\mp)$ pair, representing a small sub-class of the full two-loop diagrams. During the EWPT, the relevant CP-violating sources would be those involving the Higgsino-bino processes in Fig. \ref{fig:vev}, corresponding to \lq\lq neutralino-driven" EWBG.  A summary of the relation between baryon asymmetry and EDMs of the electron and neutron are given in Fig.~\ref{fig:mssmcpv} in the region of resonant electroweak gaugino-Higgsino EWBG. As indicated by the inner most contours, 
to conclusively test or rule out MSSM EWBG  would require improvements in the sensitivity of EDM searches by roughly two orders of magnitude, roughly consistent with the goals of the next generation of experiments. 

Going beyond the MSSM, it is possible to introduce new CPV interactions in the scalar sector that could evade present EDM constraints but still generate the CPV sources as needed for EWBG~\cite{Huber:2006wf,Huber:2006ma,Blum:2010by}. The authors of Ref.~\cite{Huber:2006wf} studied the CPV sources in the NMSSM and found that the presence of the additional gauge singlet superfield gives rise to a new CPV source that is second order in the $\epsilon_w$ expansion -- associated with a \lq\lq semi-classical force" term in kinetic theory --  and that may contribute strongly away from the resonant regime. This source depends on the same CPV phases as in the MSSM, so one must contend with constraints from EDM searches. At the time this work was completed, a minimum first and second generation sfermion mass of 1 TeV was sufficient to evade the existing $d_{e,n}$ bounds.  More recently, the authors of Ref.~\cite{Blum:2010by} observed that a new phase associated with the gauge-singlet extension of the MSSM could successfully drive EWBG through both top and stop sources as well as Higgsino-bino interactions (for a related study, see Ref.~\cite{Cheung:2012pg}). The dominant constraints on this phase are associated with two-loop contributions to the down quark \lq\lq chromo-electric dipole moment" as it might generate a $^{199}$Hg atomic EDM. Nonetheless, NMSSM EWB remains viable even with these stringent constraints. An earlier and more extensive U(1$)^\prime$ extension was studied by the authors of Ref.~\cite{Kang:2004pp}, who found a sufficiently large baryon asymmetry could be generated fromCPV sources associated with spontaneous CPV during the EWPT and whose CPV effects relaxed to sufficiently small values in the broken phase so  as to evade the EDM bounds. 

In all of the foregoing scenarios, the CPV interactions are flavor diagonal. Recently, some attention has focused on the possibility that flavor non-diagonal CPV might provide the source for EWBG while evading the one-loop EDM bounds \cite{Cline:2011mm,Tulin:2011wi,Liu:2011jh}. As with SM CKM CPV, flavor non-diagonal CPV interactions that involve the second and third generation quarks would not contribute to EDMs until at least two-loop order. On the other hand, the associated CPV effects could appear in the $B$- and/or $D$-meson system, allowing for a potential probe of this possibility. To illustrate, we consider the schematic 2DHM of Ref.~\cite{Liu:2011jh}, applied to the second and third generation fermions as needed for CPV in the $b\to s$ transitions. The corresponding interaction is
\bea
\mathcal{L} &=&   \lambda^u_{ij} \bar{Q}^i (\epsilon H_d^\dagger) u^j_R -  \lambda^d_{ij} \bar{Q}^i H_d d^j_R  \nonumber \\ &&  -  y_{ij}^u \bar{Q}^i H_u u^j_R  + y_{ij}^d \bar{Q}^i (\epsilon H_u^\dagger) d^j_R   + h.c. \ \ \ ,
\label{eq:2hdmcpv}
\eea 
where $(i,j)$ run over the second and third generations. 
For purposes of illustration, one may make several simplifying assumptions, including taking $\tan\beta=1$ at $T=0$ and setting $y_{sb}=\lambda_{sb}=0$ and retaining the one CPV phase that remains in the limit of vanishing $y_{ss}$ and $\lambda_{ss}$ (after field redefinitions): $\theta_{\lambda_{bs}}=\mathrm{Arg}(\lambda_{bs})$. In the VEV-insertion approximation the CPV sources are proportional to $|\lambda_{bs}|^2 \sin\theta_{\lambda_{bs}}$.  On the other hand, CPV in the $B_s$ system arises from tree-level exchange of the Higgs field $H_{bs}=-\cos\beta H_u+\sin\beta H_d^\dag$, leading to the mixing operator
\begin{eqnarray}
&\frac{ \zeta_{bs}^2}{\Lambda_{bs}^2} (\bar{b}_L s_R)(\bar{b}_L s_R), \ \  {\rm with}  \ \  \Lambda_{bs} \sim  m_{H_{bs}}^2 / v    \nonumber   
\label{eq:bsoperator}
\end{eqnarray}
with $\Lambda_{bs} \sim  m_{H_{bs}}^2 / v$ and $\zeta_{bs}\propto |\lambda_{bs}|(1\pm \mathrm{exp}(i\theta_{\lambda_{bs}})|$ depending on the signs of the $y_{ij}$. 

\begin{figure}[ttt]
\begin{center}
  \includegraphics[width=0.5\textwidth]{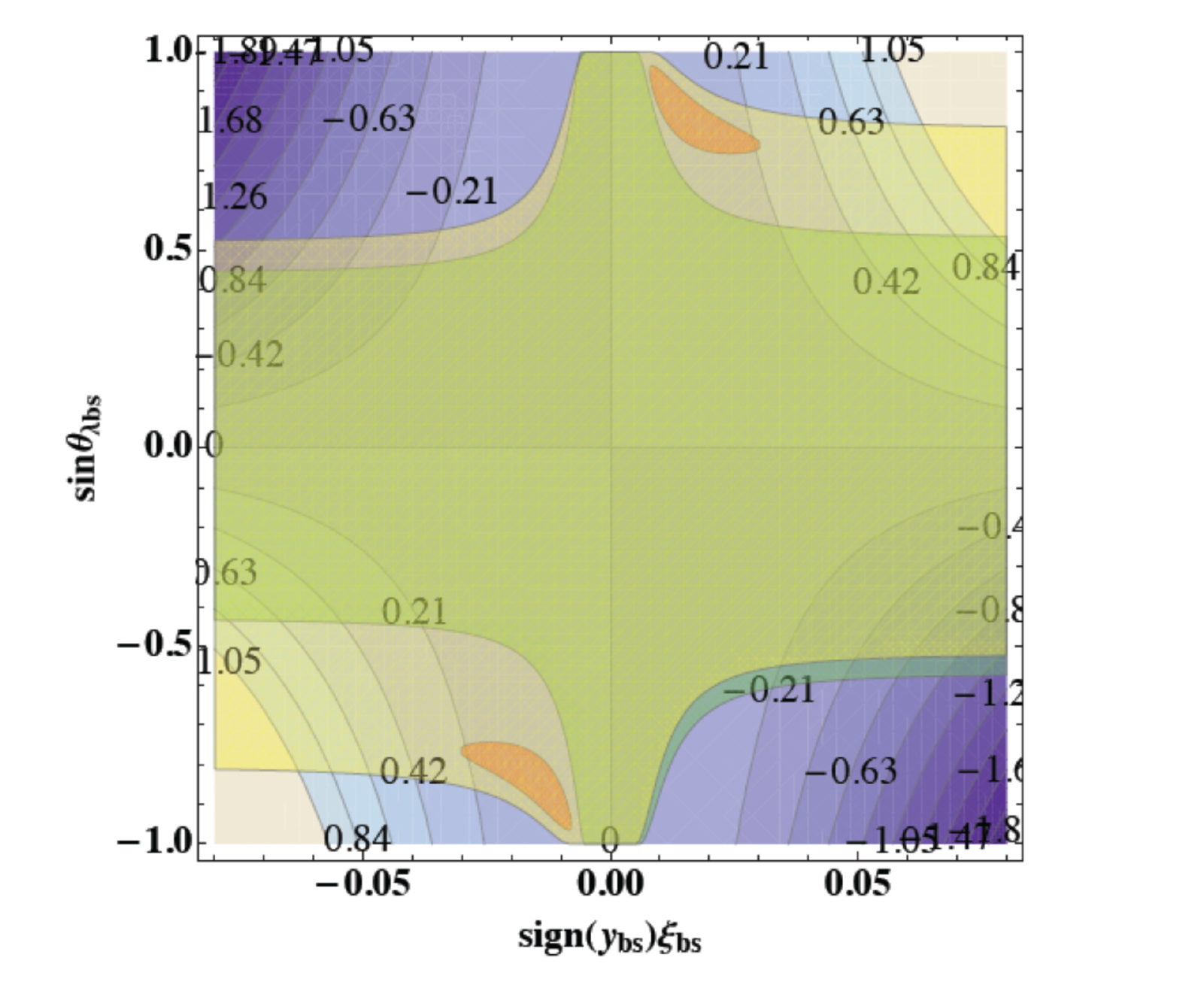}
\end{center}
\caption{\label{fig:cpvb}
Contours of constant baryon asymmetry in the \lq\lq Electroweak beautygenesis" scenario of Ref.~\cite{Liu:2011jh}.  Figure reprinted with permission from 
T. Liu, M. Ramsey-Musolf and J. Shu, 
\href{http://link.aps.org/doi/10.1103/PhysRevLett.108.221301}{Phys. Rev. Lett. 108, 221301 (2012)}, copyright 2012 by the American Physical Society.
}
\end{figure}

The resulting curves of constant baryon asymmetry, shown in the $(\sin\theta_{\lambda_{bs}}, \lambda_{bs}/\sqrt{2})$ plane are given in Fig.~\ref{fig:cpvb}, along with constraints from the Tevatron and LHCb experiments. Note that in this restricted, schematic model, non-negligible contributions to the baryon asymmetry can occur.  Future improvements in the sensitivity of LHCb studies could either uncover CPV in a region consistent with a significant contribution to the asymmetry or place stringent constraints on this possibility. On the other hand, EDM searches are relatively insensitive to the CPV phase $\theta_{\lambda_{bs}}$, as the operator in Eq.(\ref{eq:bsoperator}) contains no first generation quarks and involves only flavor non-diagonal combinations of the opposite chirality quarks. Clearly there exists considerable room for additional studies of flavor non-diagonal EWBG, including a more extensive study of the framework discussed here as well as those analyzed in Ref.~\cite{Cline:2011mm,Tulin:2011wi}. The extent to which this possibility could provide a viable EWBG alternative to flavor diagonal CPV in EWBG in light of increasingly stringent EDM limits remains to be seen.

\subsection{The High Energy Frontier: CP-Violation and Phase Transitions}
\label{sec:highenergy}

  New particles related to EWBG can potentially be created at
observable rates at high-energy colliders such as the LHC.
This is especially true of new colored states that help to strengthen 
the EWPT, such as a light stop in the MSSM or more generally 
a new $X$ scalar as discussed in Sec.~\ref{sec:ewpt}.  
To have an adequate effect, these particles must be lighter than
about $m_X \lesssim 200\,\gev$ implying very large LHC (and Tevatron)
production cross sections.  Even so, such states
be consistent with existing collider limits.  

  One way to hide a light colored scalar is to have it decay to dijets.
This can arise from a $\bar{X}q_iq_j$ coupling.  Even though the dijets
in this case would form an invariant-mass peak at the $X$ mass, 
the backgrounds at the LHC (and the Tevatron) are so large that they 
swamp the signal in the low-mass region.  Moreover, the cuts used in 
existing searches single or paired dijet resonances at both the LHC 
and the Tevatron do not extend or have limited sensitivity to lighter masses 
$m_X \lesssim 200\,\gev$~~\cite{Yu:2011cw,Aad:2011yh,cms:2012ex}.  
Some sensitivity could be recovered with more generous 
cuts~\cite{Kilic:2008pm}, or if the
$X$ decays involve heavy flavor~\cite{Choudhury:2005dg}.

\begin{figure}[ttt]
\begin{center}
  \includegraphics[width=0.5\textwidth]{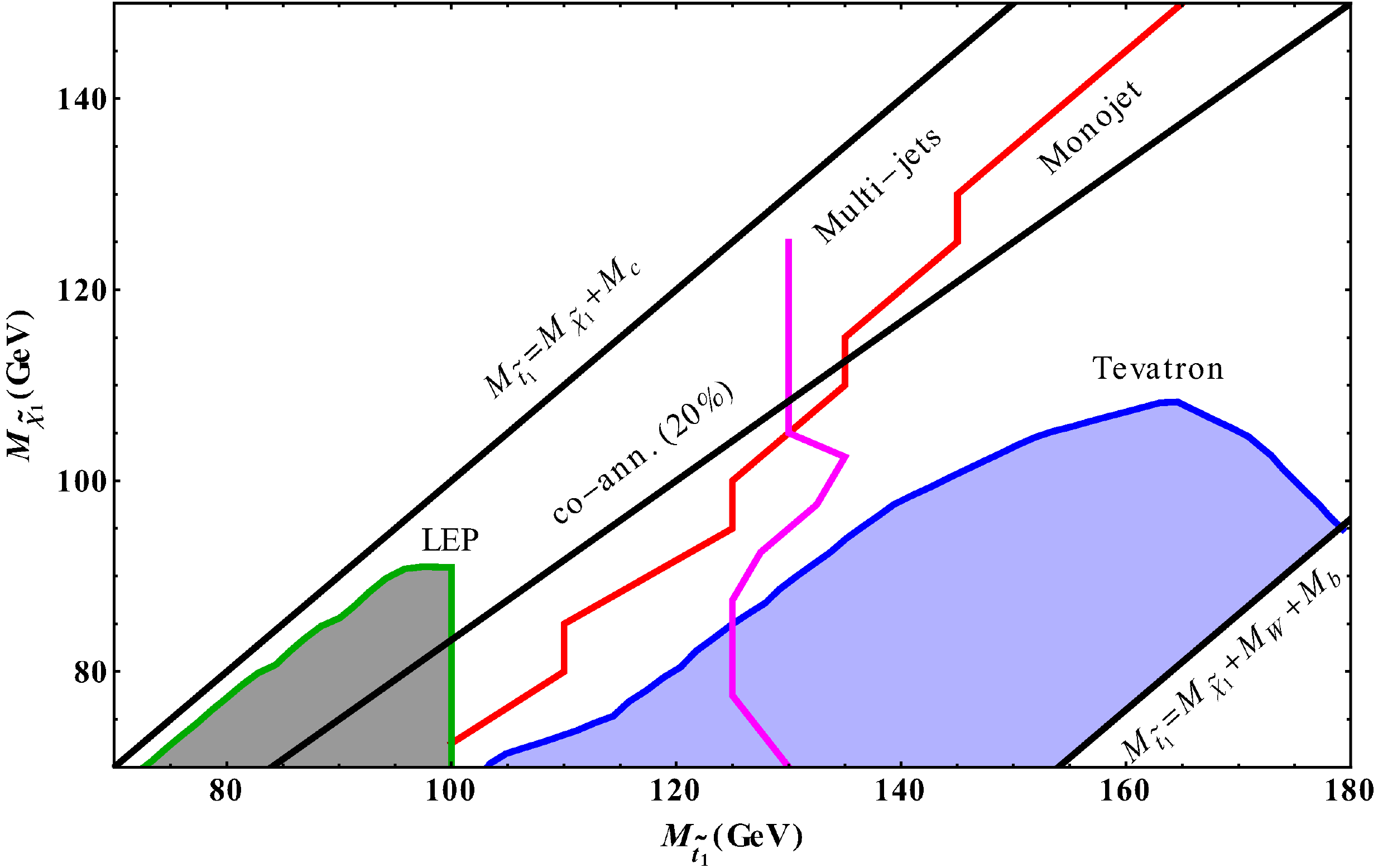}
\end{center}
\caption{\label{fig:stop-shafi}
Tevatron and LHC limits on a light stop that decays to a charm quark 
and a neutralino LSP, $\tilde{t}_1\to c\,\chi_1^0$.  Figure from
Ref.~\cite{He:2011tp}.}
\end{figure}

  A second way for a light colored state $X$ to have escaped detection 
is for it to decay to a light quark and a long-lived neutral 
$N$ fermion (that could be the dark matter).  This is natural in the MSSM, 
where the light stop ($X = \tilde{t}_1$) can decay to a charm quark and 
the lightest neutralino ($N = \chi_1^0$).
For small $X\!-\!N$ mass splittings, the decay
products are very soft and difficult to detect using the standard
searches for multiple jets and missing energy, and dedicated Tevatron
searches only limit splittings above about 
$35\,\gev$~\cite{cdfstop,Abazov:2008rc}.
In Fig.\ref{fig:stop-shafi} we illustrate the current limits on
a light stop that decays via $\tilde{t}_1\to c\chi_1^0$ based
on data from the Tevatron and the LHC.
Instead, some sensitivity can be recovered through monojet searches, 
requiring a single hard jet and missing 
energy~\cite{Carena:2008mj,Bi:2011ha,Ajaib:2011hs,He:2011tp},
as can also be seen in Fig.~\ref{fig:stop-shafi},
although a full test of the parameter space will 
be challenging~\cite{Drees:2012dd}.  
A potential further probe may be provided by the decays of stoponium,
a bound state consisting of a stop and its anti-particle~\cite{Martin:2008sv,Martin:2009dj,Younkin:2009zn,Barger:2011jt}.

  The new physics required for EWBG must also couple significantly to
the $SU(2)_L$-doublet Higgs field, and this can potentially induce observable 
changes in the production and decay properties of the Higgs boson.   
A very significant effect arises from colored $X$ scalars
that couple to the Higgs as in Eq.(\ref{eq:hportal}).  Such scalars
will contribute in loops to the amplitudes for Higgs boson production through
gluon fusion and Higgs decay to diphotons.  
Large positive values of the coupling
$Q>0$ are needed to induce a strongly first-order phase transition.
In this case their contribution to gluon fusion interferes constructively with the dominant
top quark loop in the SM, and destructively with the dominant $W^{\pm}$ loop
for diphoton decay.  The net result is a significant enhancement in the
rate of gluon fusion that is closely related to the strength of the 
electroweak phase transition, and a more modest decrease in the branching
fraction to diphotons~\cite{Cohen:2012zt}.  
Similar results are found for the light stop of the MSSM~\cite{Menon:2009mz}.

\begin{figure}[ttt]
\begin{center}
  \includegraphics[width=0.5\textwidth]{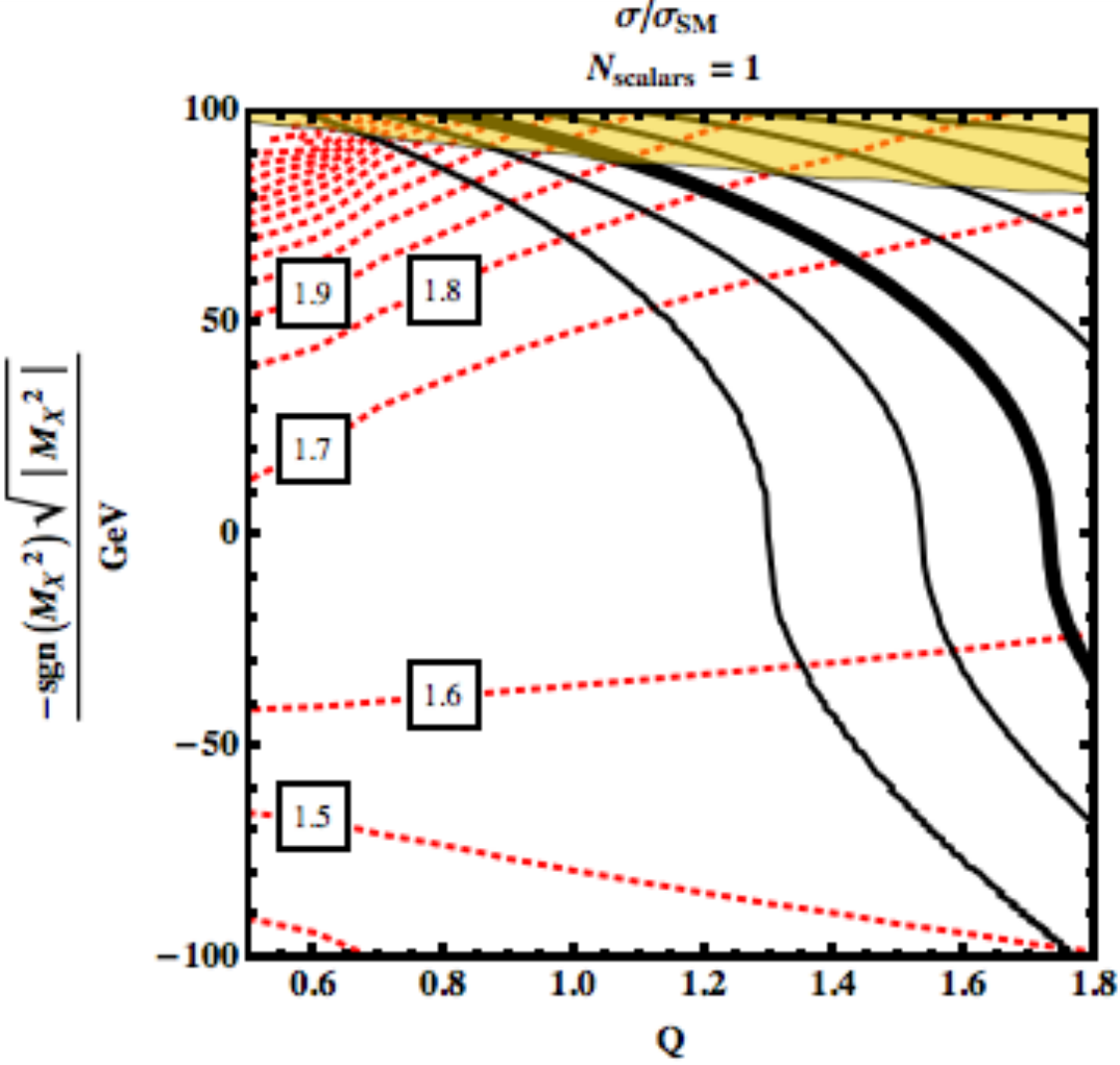}
\end{center}
\caption{\label{fig:ggtohiggs}
Rates of gluon fusion Higgs boson production relative to the SM (red dotted lines) and contours of $\phi_c/T_c$ (black solid lines) for one new color-triplet scalar for given values of the parameters $-\mathrm{sgn}\left(M_X^2\right) \sqrt{|M_X^2|}$ and $Q$.  
The thick solid corresponds to $\phi_c/T_c = 0.9$ and the adjacent solid lines show steps of $\Delta(\phi_c/T_c) = 0.2$.  Figure from Ref.~\cite{Cohen:2012zt}.}
\end{figure}

  In Fig.~\ref{fig:ggtohiggs} we show the 
enhancement of the gluon fusion rate from $X$ relative to the SM
(red dotted lines) as a function of $Q$ and the mass parameter $M_X^2$,
together with an estimate for where the phase transition is strong
enough for EWBG (to the right of the thick black solid line).
Gluon fusion is the dominant Higgs production mode at the LHC,
and is the primary production channel for the $\gamma\gamma$, 
$W^+W^-$ and $Z^0Z^0$ decay searches at the LHC.  Indeed,    
The enhancement of the gluon fusion rate implied by this mechanism
of strengthening the electroweak phase transition is already strongly
constrained by current LHC and Tevatron Higgs 
searches~\cite{Cohen:2012zt,Curtin:2012aa}.  

  New uncoloured $X$ particles coupling to the Higgs boson as 
in Eq.(\ref{eq:hportal}) can also make the phase transition more strongly 
first-order, although their effect tends to be weaker than the coloured case.
If such a state has a non-trivial electric charge, it will modify
the Higgs branching fraction to 
$h^0\to \gamma\gamma$~\cite{Cohen:2012zt,Borah:2012pu}. 
As in the coloured case,
the interference with the $W^{\pm}$ loop is destructive when the phase
transition is made more strongly first order.  The net effect, therefore,
is to decrease the Higgs branching to diphotons while leaving the
gluon fusion rate largely unchanged.  Uncoloured $X$ particles
could also potentially be probed directly by their electroweak
production channels, or their effects on the Higgs 
self-coupling~\cite{Noble:2007kk}.

  The electroweak phase transition can also be made more strongly 
first-order if there are other fields that develop VEVs in the early
Universe at about the same time as the Higgs.  A simple example of this 
is the real singlet model presented in Section~\ref{sec:ewpt}~\cite{Profumo:2007wc}.  In this case, 
there will be an additional fundamental scalar boson in the theory that will mix
with the $SU(2)_L$-doublet Higgs excitation.  The resulting real scalar 
mass eigenstates will consist of a SM-like $h_1$ and a singlet-like $h_2$.
Decays of $h_1$ are frequently similar to the SM, but can be changed
radically if $h_1\to h_2h_2$ is kinematically allowed.  
The decays of the $h_2$ state are typically inherited from $h_1$,
so the chain $h_1\to h_2h_2$ is likely to produce $4b$, $2b2\tau$,
and $4\tau$ final states, which can be distinctive but challenging
to find a hadron colliders~\cite{Carena:2007jk,Chang:2008cw}.  
On the other hand, the singlet-like $h_2$ state may have more exotic
decay channels if there are other light states in the theory,
such as a light singlet fermion~\cite{Menon:2004wv,Barger:2006dh}.
More generically, one could expect a significant signal reduction in conventional
SM Higgs search channels due to the effects of mixing and $h_1\to h_2h_2$ decays
if kinematically allowed~\cite{Profumo:2007wc}, as well as the appearance of the second state $h_2$~\cite{Barger:2007im}.

  New light particles are also needed to induce CP violation 
in the expanding bubble walls.  In many cases,
they carry non-trivial electroweak charges and couple to the
varying-Higgs background.  This is true of the MSSM, where the
main source comes from light neutralinos and charginos~\cite{Cirigliano:2006dg}.
Direct searches for such particles created via their electroweak
production modes are underway at the LHC, and some relevant exclusions
have been obtained~\cite{atlasino}.  Even so, the detailed signals depend very
sensitively on the decay channels of the new states.

\subsection{The Cosmological Frontier: Gravity Waves and More}
\label{sec:gw}

  The strongly first-order phase transition needed for EWBG can produce
a cosmological signal in the form of gravity waves\cite{Kosowsky:1992rz,Kamionkowski:1993fg}.  As discussed
above, the phase transition proceeds by the formation of bubbles
of the electroweak broken phase within the surrounding symmetric-phase
plasma.  Gravitational radiation is created by the turbulent expansion
of the walls~\cite{Kosowsky:2001xp} and their subsequent collisions as they 
coalesce~\cite{Kosowsky:1992rz,Kamionkowski:1993fg}.  The net effect of the many bubble
collisions that occurred within the current Hubble radius is a uniform
stochastic background of gravitational radiation with a characteristic spectrum.

  The spectrum and intensity of gravity waves created by a strongly
first-order phase transition depend on three parameters characterizing
the transition~\cite{Kamionkowski:1993fg,Kosowsky:2001xp}: 
the latent heat $\alpha$ released by the phase transition at the nucleation
temperature $T_n$ relative to the background radiation energy,
the characteristic rate of bubble nucleation $\beta$,
and the bubble wall velocity $v_b$.
All three quantities can be computed from the finite-temperature 
effective potential discussed in Sec.~\ref{sec:ewpt}.

  Estimates of the gravity wave signals produced by a strongly
first-order electroweak phase transition suggest that it will be 
very difficult to detect in the foreseeable 
future~\cite{Nicolis:2003tg,Grojean:2006bp,Huber:2007vva}.  
The signal
is typically too low in frequency to be picked up by the LIGO
experiment, but it may be visible at LISA if the transition
is extremely strong.  The prospects of discovery are considerably
better at BBO, but even in this case the signal from the phase
transition could be obscured by larger gravity waves signals due to 
astrophysical processes or inflation~\cite{Grojean:2006bp}.

  New physics related to EWBG can also play an important role
of other aspects of cosmology.  A specific example of this is 
dark matter~(DM) in the MSSM.  Here, the lightest neutralino
can play a key role in generating the baryon asymmetry as described
in Sec.~\ref{sec:cpbg}, and can also make up the DM if it is the
lightest superpartner~(LSP).  When the LSP is Bino-like, as motivated
by the limits from EDM searches, the relic abundance tends to be
too large.  However, MSSM EWBG also requires a very
light stop, and this state can reduce the Bino-like LSP abundance
to an acceptable level by coannihilation~\cite{Balazs:2004bu,Balazs:2004ae}.
A similar interplay between the new states contributing to EWBG
and those making up the DM is also found in variety of other models.

\section{Summary
\label{sec:conc}}

Electroweak baryogenesis remains a theoretically attractive and experimentally testable mechanism to generate the baryon asymmetry of the Universe. During the past decade, much of the theoretical attention  has focused on leptogenesis -- motivated in part by the discovery of neutrino oscillations and the relative simplicity of this scenario.  More recently, however, there has been a resurgence of  interest in EWBG  due to its testability and the advent of new probes of physics at the terascale. Searches for new particles at the LHC could discover degrees of freedom that were thermally abundant during the electroweak phase transition and whose interactions could have engendered a strong first order transition while giving rise to particle asymmetries needed to seed the production of baryons. At the same time, a generation of new searches for permanent electric dipole moments will provide ever more powerful probes of possible flavor diagonal CP-violation in these new interactions that is an essential ingredient for successful EWBG. In parallel, new mechanisms involving flavor non-diagonal terascale CP-violation are under study, and these mechanisms may have signatures in experiments involving B-mesons at the LHC or super-B factories. In short, one may anticipate that either the ingredients for successful EWBG will be  uncovered during the coming decade or that this scenario will be sufficiently constrained that more speculative, high-scale baryogenesis scenarios such as leptogenesis are left standing as the most viable alternatives. 

Achieving this confrontation of experiment with EWBG requires having in hand the most robust theoretical tools for computing the baryon asymmetry and a sufficiently broad phenomenological framework. In both respects, the past decade has witnessed considerable advances. Substantial effort has been devoted to deriving and solving the relevant set of transport equations that underlie the production of left-handed particle number asymmetries, while new attention has focused on achieving more theoretically well-defined computations of the phase transition properties. Nonetheless, there exists ample room for further improvements. In the case of the EWPT, future work could include development of additional methods for carrying out gauge-invariant perturbative computations, calculations of the fluctuation determinant in BSM scenarios, and investigation of the other sources of uncertainty that enter the BNPC of Eq.(\ref{eq:bnpc}). New Monte Carlo studies in representative BSM theories would also provide important benchmarks for gauging the validity of perturbative computations and their phenomenological implications. Continued refinements of the transport machinery would include application of the recent developments for scalar fields to fermionic systems, updated analyses of the \lq\lq semiclassical force" terms, and resolution of questions surrounding the coherence shells.  From the standpoint of phenomenology, an interesting new direction entails the possible role of flavor non-diagonal CP-violation. Further study of signatures of BSM scalar sectors, including modified Higgs production cross sections, branching ratios, and exotic final states  -- as well as scenarios in which interesting scalar sector extensions could evade discovery at the LHC - is also an obvious priority. 

More broadly speaking, explaining the origin of the visible matter of the Universe continues to be one of the primary motivations for seeking what lies beyond the Standard Model. While it is by no means certain that the explanation lies at the terascale, the time is ripe to address this possibility with vigor as part of the larger effort to determine what new symmetries and degrees of freedom -- if any -- are associated with the electroweak chapter of cosmic history. Whatever the outcome of the EWBG theory-experiment interface, one can expect this endeavor to yield important new insights into the fundamental laws of nature and their cosmological implications.

\newpage

\ack 
The authors thank V. Cirigliano, S. Profumo, and Carlos Wagner for helpful discussions and critical reading of the manuscript. This work was supported in part by U.S. Department of Energy Contract DE-FG02-08ER41531(MJRM) and the Wisconsin Alumni Research Foundation (MJRM), as well as by NSERC (DEM).  DEM and MJRM would also like to thank the \href{http://publish.aps.org/}{American Physical Society}, \href{http://www.journals.elsevier.com/nuclear-physics-b/}{Elsevier},
and the \href{http://jhep.sissa.it/jhep/}{Journal of High Energy Physics~(JHEP)} and the corresponding authors for permission to reproduce figures.

\section*{References} 

\end{document}